\documentclass[12pt, draftclsnofoot, onecolumn]{IEEEtran}
\usepackage[cmex10]{amsmath}
\usepackage{amssymb}
\usepackage{cite}
\usepackage{graphicx}
\usepackage{array,color}
\usepackage{multirow}
\usepackage{amsmath}
\usepackage{stfloats}
\usepackage{graphicx}
\usepackage{subfigure}
\usepackage{tabularx}
\usepackage{epsfig,epsf,color,balance,cite}
\usepackage{setspace}
\usepackage{bm}
\usepackage{textcomp}
\usepackage{algorithmic}
\usepackage{algorithm}

\usepackage{comment}
\usepackage{subfigure}

\usepackage{caption}
\usepackage{graphicx}
\usepackage{setspace}
\usepackage{diagbox}

\usepackage{multirow}
\usepackage{graphicx}
\makeatletter

\newcommand{\Rmnum}[1]{\expandafter\@slowromancap\romannumeral #1@}
\makeatother
\usepackage{booktabs}
\usepackage{amsmath}

\makeatother
\pdfoutput=1
\begin{document}

\title{Low-overhead  Beam Training Scheme for Extremely Large-Scale RIS in Near-field}

\author{
Wang Liu, Cunhua Pan, $\textit{Senior Member, IEEE}$, Hong Ren, Feng Shu, Shi Jin, $\textit{Senior Member, IEEE}$, and Jiangzhou Wang, $\textit{Fellow, IEEE}$ \thanks{
\emph{Corresponding author: Cunhua Pan and Hong Ren.}

Wamg Liu, Cunhua Pan, Hong Ren and Shi Jin are with National Mobile Communications Research Laboratory, Southeast University, Nanjing 210096, China. (e-mail:{wangliu, cpan, hren, jinshi}@seu.edu.cn).

Feng Shu is with the School of Electronic and Optical Engineering, Nanjing University of Science and Technology, Nanjing 210094, China, and also with the College of Physics and Information, Fuzhou University, Fuzhou 350116, China (e-mail: shufeng@njust.edu.cn).

Jiangzhou Wang is with the School of Engineering, University of Kent,
CT2 7NZ Canterbury, U.K. (e-mail: j.z.wang@kent.ac.uk).
		
 } }

\maketitle

\begin{abstract}
Extremely large-scale reconfigurable intelligent surface (XL-RIS) has recently been proposed and is recognized as a promising technology that can further enhance the capacity of communication systems and compensate for severe path loss . However, the pilot overhead of beam training in XL-RIS-assisted wireless communication systems is enormous because the near-field channel model needs to be taken into account, and the number of candidate codewords in the codebook increases dramatically accordingly. To tackle this problem, we propose two deep learning-based near-field beam training schemes in XL-RIS-assisted communication systems, where deep residual networks are employed to determine the optimal near-field RIS codeword. Specifically, we first propose a far-field beam-based beam training (FBT) scheme in which the received signals of all far-field RIS codewords are fed into the neural network to estimate the optimal near-field RIS codeword. In order to further reduce the pilot overhead, a partial near-field beam-based beam training (PNBT) scheme is proposed, where only the received signals corresponding to the partial near-field XL-RIS codewords are served as input to the neural network. Moreover, we further propose an improved PNBT scheme to enhance the performance of beam training by fully exploring the neural network's output. Finally, simulation results show that the proposed schemes outperform the existing beam training schemes and can reduce the beam sweeping overhead by approximately 95\%.

\end{abstract}

\vspace{-0.6cm}
\section{Introduction}
Recently, reconfigurable intelligent surface (RIS) has received considerable research attention and is regarded as one of the key technologies for the next-generation communication systems since it can significantly improve the energy and spectrum efficiencies of the wireless communication systems \cite{RIS_1,RIS_2,RIS_3,RIS_4}. Specifically, an RIS consists of numerous passive and controllable reflecting elements which can independently adjust the phase shift of the incident electromagnetic signal with low power consumption. By properly tuning the coefficients of the RIS reflecting elements, RIS can improve the communication signal quality and establish an additional reliable reflecting link between the user equipment (UE) and the base station (BS). Furthermore, recent advances in materials enable RIS to tune the reflecting coefficients in real time to cope with the rapidly varying wireless propagation environment \cite{RIS_material}. Reliable RIS-assisted communication systems rely heavily on accurate channel state information (CSI) for the phase shift design. However, it is challenging to obtain accurate CSI because of the passive property of the RIS.

Specifically, since the RIS lacks the ability to process the signal, only the cascade channel, i. e. the BS-RIS-UE channel, can be estimated. Accordingly, some channel estimation schemes, such as least square (LS) or minimum mean squared error (MMSE), have been proposed in \cite{LS_1,LS_2,MMSE} for cascaded channel estimation. However, the pilot overhead required for cascaded channel estimation is proportional to the number of RIS elements and becomes excessive when the number of elements is large. In order to reduce the pilot overhead, the authors of \cite{gui} proposed a novel multi-user-based channel estimation scheme, where the correlation of each user's cascade channel was explored. Furthermore, some compressive sensing (CS) based channel estimation schemes were proposed to reduce the pilot overhead, where the sparsity of the channel was leveraged \cite{CS_1,CS_2}.

Compared to channel estimation, codebook-based beam training has been widely adopted for millimetre-wave communication systems owing to its lower operation complexity \cite{RIS_bt1}. Furthermore, beam training techniques have recently been extended to RIS-assisted wireless communication systems \cite{RIS_bt1,RIS_bt2,RIS2}. Specifically, the RIS codebook is first designed based on the cascaded array steering vector. Then the RIS selects the optimal codeword from the codebook to form the phase shift based on the information acquired during the training phase. For beam training, testing all codewords in the codebook during the training phase is the most straightforward scheme; however, this will incur an excessive pilot overhead \cite{beam_sweep}. To reduce the pilot overhead, a hierarchical codebook-based beam training scheme was proposed for millimetre wave (mmWave) communication systems, where the training phase was divided into multiple parts and the range of the optimal codeword was reduced in each part \cite{h_codebok_1,h_codebok_2,h_codebok_3}. Moreover, the authors of \cite{RIS2} proposed a partial search-based beam training scheme to reduce the complexity of training.

Recently, deep learning (DL) as a branch of machine learning (ML) has received tremendous research attention and has been applied in beam training to reduce the pilot overhead since the neural networks possess a powerful ability to learn non-linear relations \cite{jindian,make1,make2,make3,qi}. For example, the authors of \cite{make3} proposed to employ a deep neural network (DNN) to determine the optimal codeword based on the information of the previous low-frequency channel. In \cite{make1}, a long and short-term memory (LSTM) network-based beam training scheme was proposed in which the received signals of the previous time slots are employed to predict the optimal beam for the next time slot. In addition, \cite{qi} demonstrated that the received signals corresponding to different codewords contain implicit non-linear relationships between them and  these relationships can be explored by the DNN to determine the optimal codeword, where only portion of the codewords need to be tested.

However, the existing beam training schemes cannot be directly applied to the RIS-assisted wireless communication systems. Specifically, the RIS is expected to employ more reflecting elements to compensate for the severe ``multiplicative fading'' effect in the cascaded channel, where the path loss of the cascaded channel is equivalent to the product of the path loss of the two links. Due to the advantages of low power consumption and low cost of the RIS, it is feasible to deploy large-scale reflecting elements at the RIS to compensate for this effect \cite{RIS_pathloss}. Consequently, extremely large-scale RIS (XL-RIS) has been proposed and regarded as one of the prospective directions of the RIS \cite{wei_codebook}. However, the increased number of reflecting elements will lead to larger Rayleigh distance, which is the boundary between the near and far fields, and the range of the near field will expand accordingly. Consequently, the near-field domain becomes non-negligible, thus scatters, as well as users, are more likely to lie in the near-field domain. For the near-field domain, the electromagnetic field structure has fundamentally changed, and the spherical wave channel model should be considered when designing the RIS codebook. Since most of the existing beam training schemes were based on far-field codebooks, it is challenging to implement them in the near-field domain. In order to address this problem, the authors of \cite{Daill_2} proposed a near-field codebook based on a spherical wave channel model at the BS, which increases distance sampling to accommodate the near-field domain. Furthermore, \cite{wei_codebook} designed a codebook for the XL-RIS, which was based on cascaded array steering vectors in the near-field domain. It can be observed from \cite{wei_codebook} and \cite{Daill_2} that the near-field codebooks contain significantly more candidate codewords than the far-field codebooks, which makes beam training in the near-field much more complicated and entails increased pilot overhead. Although the authors of \cite{wei_codebook} proposed a hierarchical near-field RIS codebook and the corresponding hierarchical beam training scheme to reduce the pilot overhead, the required pilot overhead is still excessive, and the hierarchical training scheme is highly susceptible to noise. To the best of our knowledge, the existing beam training schemes cannot address the excessive pilot overhead in the XL-RIS-assisted communication systems.

In order to fill this gap, we propose two effective deep learning-based near-field beam training schemes for XL-RIS-assisted communication systems, where the pilot overhead is significantly reduced with excellent training performance. Our contributions are summarised as follows.

\begin{enumerate}
	\item We propose a far-field beam-based beam training (FBT) scheme where a deep residual network is employed to estimate the optimal near-field RIS codeword. Specifically, the XL-RIS tests all far-field RIS codewords in the training phase and then the received signals corresponding to all far-field RIS codewords at the BS are input to the deep residual network (DRN). Next, the DRN estimates the optimal near-field RIS codeword based on the input signals by exploring the implicit relationship between the received signals of different codewords. Since the number of candidate codewords in the far-field RIS codebook is much smaller than that in the near-field XL-RIS codebook, the pilot overhead of the proposed scheme is significantly reduced.
	\item We propose a partial near-field beam-based beam training (PNBT) scheme to further reduce the pilot overhead. Compared to the FBT scheme, the scheme only tests partial near-field RIS codewords and feeds the corresponding received signals to the DRN. Furthermore, based on the PNBT scheme, we further propose an improved PNBT scheme to enhance the performance of beam training. Specifically, the improved PNBT scheme will perform additional beam test of the more likely near-field RIS codewords based on the neural network's output. 
	\item Finally, we provide numerical simulations to validate the performance of our two proposed near-field XL-RIS beam training schemes. Compared with the sweeping scheme, which exhaustively tests all codewords, the proposed scheme can obtain similar achievable rate performance, but the pilot overhead is greatly reduced. Moreover, the DRN used in this paper can better extract the features of the received signal to improve the performance of beam training than ordinary convolutional networks or fully connected networks.

\end{enumerate}

The paper is organized as follows. In Section \ref{systemmodel1}, we firstly introduce the signal model in the XL-RIS aided communication system and subsequently describe the far-field and near-field channel models. The corresponding beam training models are also presented in Section \ref{systemmodel1}. Section \ref{DL_model} first shows the problem formulation and the neural network structure. Then, two near-field RIS beam training schemes are presented. In Section \ref{simulation}, we provide the simulation results. Conclusions are drawn in  Section \ref{conclusion}.

In this paper, we adopt the following notations: Vectors and matrices are represented in bold lower case and bold upper case, respectively, e.g., $\boldsymbol{a}$ and $\boldsymbol{A}$, while $a$ and $\mathcal{A}$ denote a scale and a set. $\left ( \cdot  \right )^{\ast }$, $\left ( \cdot  \right )^{T}$ and $\left ( \cdot  \right )^{H}$ represent conjugate, transpose and conjugate transpose, respectively. $\left|\cdot  \right|$ denotes absolute value. $\left [\mathcal{A}  \right ]_{i}$ denotes the $i$-th element of $\mathcal{A}$. $\left \langle \cdot  \right \rangle$ denote the  order operation, e.g., for $\mathcal{A}=\left\{a_{1}, a_{2}, \cdots, a_{n}\right\},\langle\mathcal{A}\rangle=\left\{a_{\sigma_{1}}, a_{\sigma_{2}}, \cdots, a_{\sigma_{n}}\right\}$ with $a_{\sigma_{1}} \geq a_{\sigma_{2}} \geq \cdots \geq a_{\sigma_{n}}$. $\mathcal{C} \mathcal{N}(\mu ,\sigma )$  represents the Gaussian distribution with mean of $\mu$ and  variance of $\sigma$. $\mathcal{U}(a,b)$ represents the  uniform distribution between $a$ and $b$. Finally, diag(a) represents the diagonal matrix using vector as the diagonal elements.

\section{System Model}\label{systemmodel1}

\subsection{Signal Model}

\begin{figure}[t]
	\centering
	\includegraphics[width=3in]{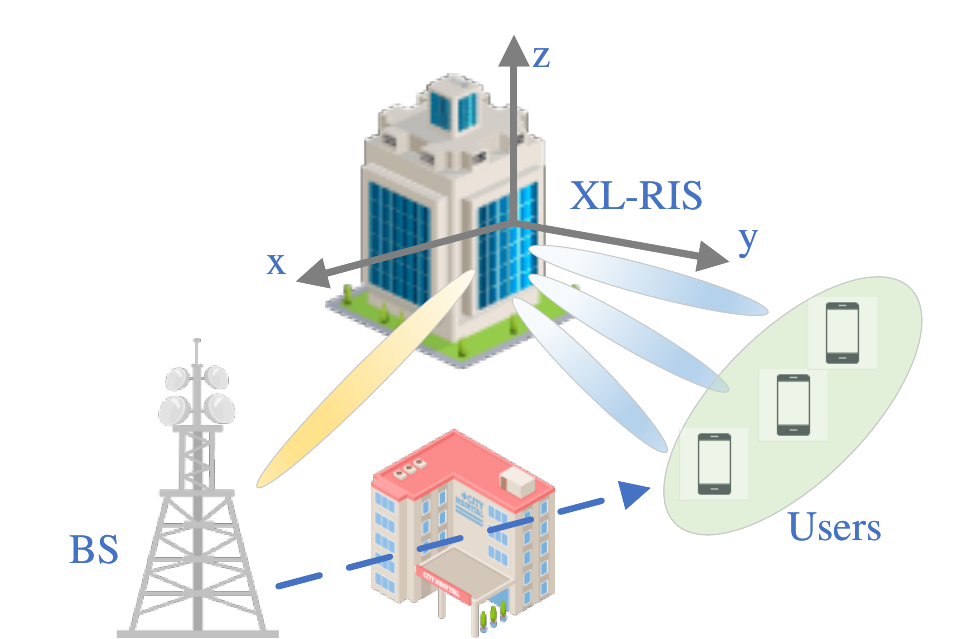}
	\caption{XL-RIS-assisted mmWave wireless communication system. }
	\label{system}\vspace{-0.9cm}
\end{figure}

As shown in Fig. \ref{system}, we consider an uplink time division duplexing (TDD) based orthogonal frequency division multiplexing (OFDM) communication system, where the direct link is blocked by obstacles, and an XL-RIS is employed between the BS and the $K$ users to assist communications\cite{ofdm}. The BS is equipped with $M$-antenna ($M=M_{1}\times M_{2}$) uniform planar array (UPA), and each user is equipped with a single antenna. The XL-RIS with $N=N_{1}\times N_{2}$ elements is deployed in the x-z plane and its centre is located at the origin of the coordinate axis. It is noted that there is a separate control link between the BS and the RIS, which is utilized by the BS to adjust the phase shifts of the XL-RIS reflection elements \cite{RIS1, RIS2, RIS3}.

Let us deonte $\textbf{G}\in \mathbb{C}^{N\times M}$ as the channel matrix from the BS to the XL-RIS, and $\textbf{h}_{k}\in \mathbb{C}^{N\times 1}$ as the channel vector from the XL-RIS to the $k$-th user. During the uplink communication, the users transmit the pilot signals to the BS via the XL-RIS, and the processed pilot signal of the $k$-th user received at the BS side can be represented as
\vspace{-0.5cm}
\begin{equation}\label{re_y}
	\setlength\abovedisplayskip{3pt}
	\setlength\belowdisplayskip{3pt}
	y_{k}=\textbf{w}_{k}^{H}\textbf{G}\mathbf{\Phi }\textbf{h}_{k}x_{k}+\textbf{w}_{k}^{H}\textbf{n}_{k},
\end{equation}
where $\textbf{w}_{k}\in \mathbb{C}^{M\times 1}$ denotes the analog combining vector for the $k$-th user at the BS; $\mathbf{\Phi }=\textrm{diag}(\phi _{1}, \phi _{2}, \cdots \,\phi _{N})$ is a diagonal matrix and each element on the diagonal represents the phase shift of the reflecting element of the XL-RIS; $x_{k}\in \mathbb{C}$ denotes the pilot signal transmitted by the $k$-th user, which satisfies $\left|x_{k} \right|^{2}=1$; $\textbf{n}_{k}\in \mathbb{C}^{M\times 1}$ is the noise vector at the BS, and each element follows the complex Gaussian distribution of zero mean and variance of $\sigma^{2} $.

\vspace{-0.5cm}
\subsection{Channel Model}

The conversion of RIS to XL-RIS means not only the increase of the number of the reflecting elements but also the fundamental transformation of channel model and electromagnetic field structure \cite{Daill_2}. Specifically, the electromagnetic radiation field consists of two main regions, i.e., the far-field and near-field regions, and these two regions are divided by Rayleigh distance $Z=\frac{2D^{2}}{\lambda _{c}}$, where $D$ and $\lambda _{c}$ represent the array aperture and carrier wavelength, respectively \cite{antenna_sapcing}. In conventional RIS-aided communication systems, the array aperture of the RIS is not very large, and the corresponding Rayleigh distance is small, which results in a small near-field range, and most of the scatters and targets are in the far-field domain \cite{near_field1}. However, with the increase in the array aperture of the RIS, the range of the near-field also increases, which makes the spherical wave channel model no longer neglectable. In the following, we briefly introduce the channel models in the far-field and near-field domains, respectively.

$1)$ $\textit{Far-Field Channel Model}$: In the far-field domain, the Saleh-Valenzuela channel model widely used. Specifically, the channel $\textbf{G}$ can be represented as 
\begin{equation}\label{far-g}
	\setlength\abovedisplayskip{3pt}
	\setlength\belowdisplayskip{3pt}
	\begin{split}
		\textbf{G}=\sqrt{\frac{MN}{L_{G}}}\sum_{l_{1}=1}^{L_{G}}\alpha ^{G}_{l_{1}}\mathbf{a}\left ( \vartheta ^{G_{t}}_{l_{1}} ,\psi  ^{G_{t}} _{l_{1}}\right )\mathbf{b}^{H}\left( \vartheta ^{G_{r}} _{l_{1}} ,\psi ^{G_{r}}  _{l_{1}}\right ),
	\end{split}
\end{equation}
where $L_{G}$ denotes the numbers of paths between the BS and  the XL-RIS; $\alpha ^{G}_{l_{1}}$ represents the complex path gain of the $l_{1}$-th path ; $\vartheta ^{G_{t}}_{l_{1}}(\psi  ^{G_{t}} _{l_{1}})$ and $\vartheta ^{G_{r}} _{l_{1}} (\psi ^{G_{r}}  _{l_{1}})$ denote the azimuth (elevation) angle at the BS and the azimuth (elevation) angle at the XL-RIS of the $l_{1}$-th path. In particular, the first channel path, i.e. $l_{1}=1$, represents the strongest channel path, which is generally the line-of-sight (LOS) path. 

Similarly, the channel between the XL-RIS and the $k$-th user $\textbf{h}_{k}$ can be formulated as
\begin{equation}\label{far_h}
	\setlength\abovedisplayskip{1pt}
	\setlength\belowdisplayskip{3pt}
\textbf{h}_{k}=\sqrt{\frac{N}{L_{k}}}\sum_{l_{2}=1}^{L_{k}}\alpha ^{k}_{l_{2}}\mathbf{b}\left ( \vartheta ^{k}_{l_{2}} ,\psi  ^{k} _{l_{2}}\right ),
\end{equation}
where $L_{k}$ denotes the number of paths between the XL-RIS and the $k$-th user; $\alpha ^{k}_{l_{2}}$ and $\vartheta ^{k}_{l_{2}} (\psi  ^{k} _{l_{2}})$ represents the complex path gain and the azimuth (elevation) angle of the $l_{2}$-th path, respectively. In a similar way, $l_{2}=1$ represents the strongest path.
Furthermore, $\mathbf{a}\left ( \vartheta ,\psi  \right )$ and $\mathbf{b}\left ( \vartheta ,\psi  \right )$ denote the far-field array steering vectors for the BS and the RIS, respectively. Specifically, $\mathbf{a}\left ( \vartheta ,\psi  \right )$ can be formulated as \cite{steer_vector}
\begin{equation}\label{far_steer}
	\setlength\abovedisplayskip{3pt}
	\setlength\belowdisplayskip{3pt}
	\mathbf{a}(\vartheta, \psi)=\frac{1}{\sqrt{M}}\left[e^{-j 2 \pi d \cos (\psi) \mathbf{m}_1 / \lambda}\right] \otimes\left[e^{-j 2 \pi d \sin (\psi) \cos (\vartheta) \mathbf{m}_2 / \lambda}\right],
\end{equation}
where $ \lambda$ and $d$ denote the wavelength and the antenna spacing, which usually satisfy $d=\frac{\lambda }{2}$; $\mathbf{m}_1=\left [ 0,1,\cdots ,M_1-1 \right ]^{T}$ and $\mathbf{m}_2=\left [ 0,1,\cdots ,M_2-1 \right ]^{T}$.

It should be noted that in mmWave communications, the received signal power is concentrated on the strongest path, and the purpose of beam training is to align the beam to the strongest path to maximize the beamforming gain while the other paths can be treated as the noise \cite{strong_path1,strong_path2}. Therefore, in this paper, we mainly consider the strongest path. Specifically, the received signal in (\ref{re_y}) can be rewritten as 
\begin{equation}\label{far_y_1}
	\setlength\abovedisplayskip{3pt}
	\setlength\belowdisplayskip{3pt}
	y_{k}=\beta _{k}\textbf{w}_{k}^{H}\mathbf{a}\left ( \vartheta ^{G_{t}}_{1} ,\psi  ^{G_{t}} _{1}\right )\mathbf{b}^{H}\left( \vartheta ^{G_{r}} _{1} ,\psi  ^{G_{r}} _{1}\right )\mathbf{\Phi }\mathbf{b}\left ( \vartheta ^{k}_{1} ,\psi  ^{k} _{1}\right )x_{k}+n_{e,k},
\end{equation}
where $\beta _{k}=\sqrt{\frac{MN^{2}}{L_{G}L_{k}}}\alpha ^{G}_{1}\alpha ^{k}_{1}$ denotes the effective path gain of the $k$-th user; $n_{e,k}$ denotes the effective noise, which consists of the combined channel noise $\textbf{n}_{k}$ and combined received signals from the other paths. Since the locations of the BS and the RIS are generally fixed and the channel $\textbf{G}$ is essentially constant over multiple channel coherence time intervals, we assume that the analog combining vector  $\textbf{w}_{k}$ at the BS has been designed to align to the strongest path, i.e. $\textbf{w}_{k}=\mathbf{a}\left ( \vartheta ^{G_{t}}_{1} ,\psi  ^{G_{t}} _{1}\right )$. The received signal can be further formulated as \cite{wei_codebook}
\begin{equation}\label{far_y_2}
	\setlength\abovedisplayskip{3pt}
	\setlength\belowdisplayskip{3pt}
	\begin{aligned}
		y_{k} &=\beta _{k}\mathbf{b}^{H}\left( \vartheta ^{G_{r}} _{1} ,\psi  ^{G_{r}} _{1}\right )\mathbf{\Phi }\mathbf{b}\left ( \vartheta ^{k}_{1} ,\psi  ^{k} _{1}\right )x_{k}+n_{e,k}\\
		&=\beta _{k}\mathbf{b}^{H}\left( \vartheta ^{G_{r}} _{1} ,\psi  ^{G_{r}} _{1}\right )\textrm{diag}(\boldsymbol{\phi })\mathbf{b}\left ( \vartheta ^{k}_{1} ,\psi  ^{k} _{1}\right )x_{k}+n_{e,k}\\
		&=\beta _{k}\boldsymbol{\phi }^{T}\textrm{diag}\left ( \mathbf{b}^{\ast }\left( \vartheta ^{G_{r}} _{1} ,\psi  ^{G_{r}} _{1}\right) \right )\mathbf{b}\left ( \vartheta ^{k}_{1} ,\psi  ^{k} _{1}\right )x_{k}+n_{e,k}\\
		&=\beta _{k}\boldsymbol{\phi }^{T}\mathbf{b}\left ( \vartheta ^{k}_{1}-\vartheta ^{G_{r}} _{1},\psi  ^{k} _{1}-\psi  ^{G_{r}} _{1}\right )x_{k}+n_{e,k},
	\end{aligned}
\end{equation}
where $\boldsymbol{\phi }=[\phi _{1}, \phi _{2}, \cdots ,\phi _{N}]^{T}$ denotes the phase shift vector of the XL-RIS.

$2)$ $\textit{Near-Field Channel Model}$: In the near-field domain, the spherical wave channel model used \cite{near_field_G1,near_field_G2}, which can be represented as 
\begin{equation}\label{near-g}
	\setlength\abovedisplayskip{3pt}
	\setlength\belowdisplayskip{3pt}
	\begin{split}
	\textbf{G}=\sqrt{\frac{MN}{L_{G}}}\sum_{l_{1}=1}^{L_{G}}\alpha ^{G}_{l_{1}}\mathbf{c}_{t}\left ( x^{G}_{l_{1}},y^{G}_{l_{1}},z^{G}_{l_{1}}\right )\mathbf{c}_{r}^{H}\left( x^{G}_{l_{1}},y^{G}_{l_{1}},z^{G}_{l_{1}}\right ),
	\end{split}
\end{equation}
where $\left ( x^{G}_{l_{1}},y^{G}_{l_{1}},z^{G}_{l_{1}}\right )$ deontes the coordinate of the scatter on the $l_{1}$th path between the XL-RIS and the BS. In particular, $\left ( x^{G}_{1},y^{G}_{1},z^{G}_{1}\right )$ represents the coordinate of the scatter on the strongest path. Similar to the array steering vectors in the far-field domain, $\mathbf{c}_{t}\left ( x,y,z\right )$ and $\mathbf{c}_{r}\left ( x,y,z\right )$ represent the near-field steering vectors at the XL-RIS and at the BS, respectively. Take $\mathbf{c}_{r}\left ( x,y,z\right )$ as an example, the near-field steering vector can be formulated as
\begin{equation}\label{near_steer}
	\setlength\abovedisplayskip{3pt}
	\setlength\belowdisplayskip{3pt}
	\begin{split}
		\mathbf{c}_{r}\left ( x,y,z\right )=\frac{1}{\sqrt{N}}\left [ e^{-j 2 \pi \frac{d}{\lambda }D_{1,1}}, \cdots ,e^{-j 2 \pi \frac{d}{\lambda }D_{1,N_{2}}},\cdots ,\right.
		\left.e^{-j 2 \pi \frac{d}{\lambda }D_{N_{1},1}},\cdots ,e^{-j 2 \pi \frac{d}{\lambda }D_{N_{1},N_{2}}}\right ]^{T},
	\end{split}
\end{equation} 
where $D_{n_{1},n_{2}}=\sqrt{(x_{n_{1},n_{2}}-x)^{2}+y^{2}+(z_{n_{1},n_{2}}-z)^{2}}$ represents the distance from the element in the $n_{1}$-th row and $n_{2}$-th column on the XL-RIS to the scatter. $(x_{n_{1},n_{2}},0,z_{n_{1},n_{2}})$ represents the coordinate of the corresponding XL-RIS element.

Similarly, the channel between the XL-RIS and the $k$-th user $\textbf{h}_{k}$ can be formulated as
\begin{equation}\label{near_h}
	\setlength\abovedisplayskip{3pt}
	\setlength\belowdisplayskip{3pt}
\textbf{h}_{k}=\sqrt{\frac{N}{L_{r,k}}}\sum_{l_{2}=1}^{L_{k}}\alpha ^{k}_{l_{2}}\mathbf{c}_{r}\left (  x^{k}_{l_{2}},y^{k}_{l_{2}},z^{k}_{l_{2}}\right ),
\end{equation}
where $\left (  x^{k}_{l_{2}},y^{k}_{l_{2}},z^{k}_{l_{2}}\right )$ denotes the coordinate of the scatter between the XL-RIS and the $k$-th user of the $l_{2}$-th path and $\left (  x^{k}_{1},y^{k}_{1},z^{k}_{1}\right )$ denotes the coordinate of the scatter or the user on the strongest path.

Similar to (\ref{far_y_1}) and (\ref{far_y_2}), the received signal in the near-field domain can be rewritten as 
\begin{equation}\label{near_y_1}
	\setlength\abovedisplayskip{3pt}
	\setlength\belowdisplayskip{3pt}
	\begin{aligned}
		y_{k}&=\beta _{k}\mathbf{c}_{r}^{H}\left( x^{G}_{1},y^{G}_{1},z^{G}_{1}\right )\textrm{diag}(\boldsymbol{\phi })\mathbf{c}_{r}\left (  x^{k}_{1},y^{k}_{1},z^{k}_{1}\right )x_{k}+n_{e,k}\\
		&=\beta _{k}\boldsymbol{\phi }^{T}\textrm{diag}\left ( \mathbf{c}_{r}^{\ast }\left( x^{G}_{1},y^{G}_{1},z^{G}_{1}\right ) \right )\mathbf{c}_{r}\left (  x^{k}_{1},y^{k}_{1},z^{k}_{1}\right )x_{k}+n_{e,k}\\
		&=\beta _{k}\boldsymbol{\phi }^{T}\bar{\mathbf{c}}_{r}\left( (x^{G}_{1},y^{G}_{1},z^{G}_{1}),(x^{k}_{1},y^{k}_{1},z^{k}_{1})\right )x_{k}+n_{e,k},
	\end{aligned}
\end{equation}
where $\bar{\mathbf{c}}_{r}\left( (x^{G}_{1},y^{G}_{1},z^{G}_{1}),(x^{k}_{1},y^{k}_{1},z^{k}_{1})\right )$ denotes the effective near-field steering vector and can be formulated as
\begin{equation}\label{eff_near_steer}
	\setlength\abovedisplayskip{3pt}
	\setlength\belowdisplayskip{3pt}
	\begin{aligned}
		&\bar{\mathbf{c}}_{r}\left( (x^{G}_{1},y^{G}_{1},z^{G}_{1}),(x^{k}_{1},y^{k}_{1},z^{k}_{1})\right)\\
		&=\frac{1}{\sqrt{N}} [ e^{-j 2 \pi \frac{d}{\lambda }(D^{k}_{1,1}-D^{G}_{1,1})}, \cdots ,e^{-j 2 \pi \frac{d}{\lambda }(D^{k}_{1,N_{2}}-D^{G}_{1,N_{2}})},\\
		&\cdots ,e^{-j 2 \pi \frac{d}{\lambda }(D^{k}_{N_{1},1}-D^{G}_{N_{1},1})},\cdots ,e^{-j 2 \pi \frac{d}{\lambda }(D^{k}_{N_{1},N_{2}}-D^{G}_{N_{1},N_{2}})}],
	\end{aligned}
\vspace{-0.5cm}
\end{equation} 
where $D^{k}_{n_{1},n_{2}}$ and $D^{G}_{n_{1},n_{2}}$ denote the distances from the element on the XL-RIS to the scatter $(x^{k}_{1},y^{k}_{1},z^{k}_{1})$ and to the scatter $(x^{G}_{1},y^{G}_{1},z^{G}_{1})$, respectively. 

Since the propagation environment and scatter locations between the XL-RIS and the BS are generally fixed, and the channel $\textbf{G}$ remains constant over long channel coherence time, we assume that the coordinates of the scatters on the strongest path between the XL-RIS and the BS, i.e. $(x^{G}_{1},y^{G}_{1},z^{G}_{1})$, are fixed and known. Therefore, for simplicity, the received signal can be further rewritten as
\begin{equation}\label{near_y_2}
	\setlength\abovedisplayskip{3pt}
	\setlength\belowdisplayskip{3pt}
		y_{k}=\beta _{k}\boldsymbol{\phi }^{T}\bar{\mathbf{c}}_{r}\left( x^{k}_{1},y^{k}_{1},z^{k}_{1}\right )x_{k}+n_{e,k}.
		\vspace{-0.5cm}
\end{equation}

\vspace{-0.5cm}
\subsection{Beam Training Model}
It is assumed that the phase shift vector $\boldsymbol{\phi }=[\phi _{1}, \phi _{2}, \cdots ,\phi _{N}]$ of the XL-RIS  is selected from the predefined codebook, where each codeword in the codebook can form a passive beam. Based on the far-field steering vector in (\ref{far_steer}) and the received signal in (\ref{far_y_2}), the existing codebooks  for the RIS in the far-field domain are generally designed as \cite{wei_codebook}
\begin{equation}\label{far_codebook}
	\setlength\abovedisplayskip{1pt}
	\setlength\belowdisplayskip{3pt}
	\begin{split}
	\mathcal{F}=\left\{\mathbf{b}\left(\theta_{1}, \varphi _{1}\right), \ldots, \mathbf{b}\left(\theta_{1}, \varphi _{N_{2}}\right), \ldots,\right. \left.\mathbf{b}\left(\theta_{N_{1}}, \varphi _{1}\right), \ldots, \mathbf{b}\left(\theta_{N_{1}}, \varphi _{N_{2}}\right)\right\}^{\ast },
	\end{split}
\end{equation}
where $\theta_n=\frac{2 n-N_1-1}{N_1}$ with $n=1,2, \cdots, N_1$ and $\varphi_n=\frac{2 n-N_2-1}{N_2}$ with $n=1,2, \cdots, N_2$. Each element of $\mathcal{F}$ is a candidate beam codeword for the RIS in the far-field domain. Such an angular domain-based far-field codebook can exploit the far-field path's full angular information. 

However, the array steering vector in the near-field domain depends on the coordinates of the scatters, which makes the angle-only far-field codebooks no longer applicable to the channels in the near-field. For the application of the near-field domain, the design of the codebook for the XL-RIS should be based on the near-field array steering vector in (\ref{eff_near_steer}) and the received signal model in (\ref{near_y_2}).
Furthermore, the most strong path from the XL-RIS to the user is generally the  LOS path, where  $\left (  x^{k}_{1},y^{k}_{1},z^{k}_{1}\right )$ denotes the coordinate of the user, and the height of the user is generally constant. This information can be obtained at the beginning of the communication using some localization algorithms or sweeping schemes \cite{wei_codebook}. Therefore, for the design of the near-field codebooks, we mainly consider the exploration of $x^{k}_{1}$ and $y^{k}_{1}$. Specifically, the near-field codebook for the XL-RIS is designed as \cite{wei_codebook}
\begin{equation}\label{near_codebook}
	\setlength\abovedisplayskip{1pt}
	\setlength\belowdisplayskip{3pt}
	\begin{split}
		\mathcal{W}=\left\{\bar{\mathbf{c}}_{r}\left( x_{1},y_{1}\right ), \ldots, \bar{\mathbf{c}}_{r}\left( x_{S_{x}},y_{1}\right ), \ldots, \right. \left.\bar{\mathbf{c}}_{r}\left( x_{1},y_{S_{y}}\right ),\cdots ,\bar{\mathbf{c}}_{r}\left( x_{S_{x}},y_{S_{y}}\right )\right\}^{\ast },
	\end{split}
\end{equation}
where $S_{x}$ and  $S_{y} $ denote the numbers of points sampled on the x-axis and y-axis, respectively;  $x_{s_{x}}$ and  $y_{s_{y}}$  represent the sampled points on the x-axis and  y-axis  respectively. In addition, the set of all sampled points can be represented as
\vspace{-0.6cm}
\begin{equation}
	\begin{aligned}
		\Xi =\left\{\left( x_{s_{x}},y_{s_{y}}\right ) \mid\right. x_{s_{x}} =x_{\min}+(s_{x}-1)\Delta _{x}, 	y_{s_{y}} \left.=y_{\min}+(s_{y}-1)\Delta _{y} \right\},
	\end{aligned}
\vspace{-0.5cm}
\end{equation}
where $x_{\min}$ and $y_{\min}$ denote the minimum value of sampling points; $\Delta _{x} $ and $\Delta _{y} $ denote the sampling intervals on x-axis and y-axis, respectively.

Based on the well-designed codebook, the beam training aims to search for the optimal codeword that enables the system to achieve the maximum achievable rate. In order to search for the optimal codeword, the users need to send the pilot signals to the BS via the XL-RIS over $Q$ time slots, and the XL-RIS will select different codewords from the codebook to form the passive beam at each time slot. Based on the received signal in (\ref{near_y_2}), the pilot signal of the $k$-th user received at the BS at the $q$-th time slot can be expressed as
\begin{equation}\label{re_y_q}
	\setlength\abovedisplayskip{2pt}
	\setlength\belowdisplayskip{2pt}
	y_{k,q}=\beta _{k}\boldsymbol{\phi }_{q}^{T}\bar{\mathbf{c}}_{r}\left( x^{k}_{1},y^{k}_{1},z^{k}_{1}\right )x_{k,q}+n_{e,k,q},
\end{equation}
where $x_{k,q}$ denotes the pilot signal sent by the $k$-th user to the BS at the $q$-th time slot; $\boldsymbol{\phi }_{q}$ and  $n_{e,k,q}$ represent the phase shift vector of the XL-RIS and effective channel noise at the $q$-th time slot, respectively. 

After $Q$ time slots, the BS can obtain a $Q$-dimensional vector of received signals. By assuming $x_{k,q}=1$, the vector of received signals $\textbf{y}_{k}=\left [y_{k,1} ,y_{k,2},\cdots ,y_{k,Q} \right ]^{T}$ can be expressed as
\begin{equation}\label{re_y_Q}
	\setlength\abovedisplayskip{2pt}
	\setlength\belowdisplayskip{2pt}
\textbf{y}_{k}=\beta _{k}\mathbf{\Theta }\bar{\mathbf{c}}_{r}\left( x^{k}_{1},y^{k}_{1},z^{k}_{1}\right )+\textbf{n}_{e,k},
\end{equation}
where $\mathbf{\Theta }=\left [ \boldsymbol{\phi }_{1}^{T} ,\boldsymbol{\phi }_{2}^{T},\cdots ,\boldsymbol{\phi }_{Q}^{T}\right ]^{T}$ and $\textbf{n}_{e,k}=\left [n_{e,k,1},n_{e,k,2} ,\cdots ,n_{e,k,Q} \right ]^{T}$.

Furthermore, the users are assumed to adopt orthogonal frequency subcarriers \cite{wei_beamspace}, which enables the beam training between users to be independent of each other. For simplicity, the index $k$ in (\ref{re_y_Q}) can be omitted, and the received signal vector $\textbf{y}_{k}$ can be further formulated as
\begin{equation}\label{re_y_non_k}
	\setlength\abovedisplayskip{2pt}
	\setlength\belowdisplayskip{2pt}
	\textbf{y}=\beta \mathbf{\Theta }\bar{\mathbf{c}}_{r}\left( x_{1},y_{1},z_{1}\right )+\textbf{n}_{e}.
\end{equation}

Based on the received signal vector in (\ref{re_y_non_k})  and codebook in (\ref{near_codebook}), the beam training problem in the near-field domain can be formulated as
\begin{equation}\label{prob_for}
	\setlength\abovedisplayskip{3pt}
	\setlength\belowdisplayskip{3pt}
	\boldsymbol{\phi }^{\star }=\mathop{\arg\max}\limits_{\boldsymbol{\phi } \in \mathcal{W}^{N}}  \log _{2}\left(1+\frac{\left|\boldsymbol{\phi }^{T} \bar{\mathbf{c}}_{r}\left( x_{1},y_{1},z_{1}\right ) \right|^{2}}{\sigma^{2}}\right).
\end{equation}

In practice, a straightforward approach to the problem in (\ref{prob_for}) is the sweeping scheme, i.e. testing all the codewords in the codebook. Although the sweeping scheme can achieve higher achievable rates, the pilot overhead associated with the sweeping scheme is unacceptable. Specifically, for the beam sweeping scheme, the number of time slots for beam training, $Q$, is too large, which results in less time slots for transmitting useful information in each channel coherence time interval. It is even worse in the near-field domain since the near-field codebook possesses more candidate codewords than the far-field codebook \cite{wei_codebook}. 

In order to improve the efficiency of beam training, some beam training methods based on hierarchical codebooks have been proposed in the far-field domain \cite{h_codebok_1,h_codebok_2,h_codebok_3}. Similarly, a recent hierarchical training scheme based on the near-field codebook was proposed to solve the problem of excessive pilot overhead in the near-field domain \cite{wei_codebook}. The hierarchical training schemes generally contain two types of codebooks, namely wide beam codebook and narrow beam codebook, where each wide beam can cover several narrow beams. During the hierarchical training procedure, all wide beam codewords will be tested in the first stage to find the optimal wide beam, and all narrow beams covered by the optimal wide beam will then be tested in the second stage to search for the final optimal narrow beam codeword. Compared with the exhaustive sweep search, the hierarchical search schemes significantly reduce the pilot overhead.

\vspace{-0.3cm}
\section{Deep Learning for Beam Training}\label{DL_model}
\subsection{Problem Formulation}

Although the hierarchical training scheme reduces the pilot overhead required for beam training to some extent, the hierarchical training scheme is susceptible to noise as well as multipath, and the pilot overhead is high, especially in the near-field domain. From the research work in \cite{qi} and \cite{make1}, we know that the received pilot signals corresponding to different beam codewords contain implicit relationships with the optimal codeword, which provides additional information about the optimal codeword. By exploring these implicit relationships, we can reduce the pilot overhead and estimate the optimal codeword based on a small number of pilots. Therefore, in the near-field beam training, we propose to utilize the implicit relationship among codewords to estimate the optimal near-field codeword, in which only a small number of codewords need to be tested. Mathematically, the beam training model can be formulated as
\begin{equation}\label{model_1}
	\setlength\abovedisplayskip{2pt}
	\setlength\belowdisplayskip{2pt}
	s^{\star}=f\left(\textbf{y}\right), s^{\star} \in\left\{1,2, \cdots, s,\cdots ,S_{x}S_{y}\right\},
\end{equation}
where $f(\cdot )$ denotes a mapping function, $s^{\ast}$ denotes the index of the optimal codeword in the near-field codebook. 

However, the conventional estimation methods are difficult for dealing with these implicit relationships for two reasons. On the one hand, the relationships between the received signals $\textbf{y}$ and the optimal codeword are highly non-linear. On the other hand, the distribution of the equivalent noise $\textbf{n}_{e}$ is difficult to deal with, which will seriously impact the estimation performance \cite{qi,make1}. Motivated by the application of deep learning methods, we propose to employ the neural network, which has a robust ability to learn non-linear relationships and deal with noise, to address this complex estimation problem. Specifically, we train two separate neural networks, i. e. an x-axis network and a y-axis network, to estimate the x-axis index $s_{x}$ and the y-axis index $s_{y}$ of the optimal near-field codeword, respectively. Therefore, the proposed deep learning-based beam training model can be further represented as
\begin{equation}\label{model_2}
	\setlength\abovedisplayskip{3pt}
	\setlength\belowdisplayskip{3pt}
	\begin{gathered}
		s_{x}^{\star }=\mathit{f}_{x}\left ( \textbf{y} \right ),s_{x}^{\star }\in \left\{1,2,\cdots ,S_{x} \right\},\\
		s_{y}^{\star }=\mathit{f}_{y}\left ( \textbf{y} \right ),s_{y}^{\star }\in \left\{1,2,\cdots ,S_{y} \right\},
	\end{gathered}
\end{equation}
where $\mathit{f}_{x}$ and $\mathit{f}_{y}$ denote the mapping functions of the x-axis network and the y-axis network, respectively; $s_{x}^{\star }$ and $s_{y}^{\star }$ denote the x-axis index and the y-axis index of the optimal codeword in the near-field codebook. According to (\ref{model_2}), the index of the optimal near-field codeword can be represented as
\begin{equation}\label{singal_w_set}
	\setlength\abovedisplayskip{1pt}
	\setlength\belowdisplayskip{1pt}
	s^{\star }=\left ( s_{y}^{\star }-1 \right )S_{x}+s_{x}^{\star }.
\end{equation}
\vspace{-1cm}

Furthermore, for the selection of the codewords to be tested, we propose to employ all far-field codewords in (\ref{far_codebook}) and partial near-field codewords in (\ref{near_codebook}), respectively. Accordingly, we propose two different beam training schemes, namely the FBT scheme and the PNBT scheme. For the FBT scheme, all far-field beam codewords in (\ref{far_codebook}) are tested, and the corresponding pilot received signals are used as the input to the neural network to estimate the optimal near-field codeword. Since far-field codebooks have far fewer codewords than near-field codebooks, the pilot overhead is effectively reduced compared to the sweeping and hierarchical schemes in the near-field domain. By contrast, for the PNBT scheme, we select codewords from the near-field codebook at equal intervals for testing, which can also effectively reduce the pilot overhead.

It should be noted that the proposed deep learning-based beam training scheme consists of two phases, i.e. the neural network training phase and the estimation phase. For the training phase, training data containing the received pilot signal vectors and the corresponding optimal codeword index labels are collected to train the neural network, where the index labels of the optimal codewords can be obtained by conventional beam training methods. The specific neural network training steps are as follows:
\begin{enumerate}
	\item For each user, the XL-RIS tests all near-field codewords and the BS receives the corresponding received signals.
	\item The BS measures the quality of all received signals and obtains the optimal near-field codeword.
	\item A dataset is constructed by storing received signals as features and their optimal codeword as labels. Repeat and obtain multiple sets of such data.
	\item The obtained datasets are utilized to train the neural network.
\end{enumerate}

 For the estimation phase, the neural network has been fully trained and utilizes the pilot received signal vector as the input to estimate the index of the optimal codeword.

\begin{figure*}[t]
	\centering
	\subfigure["Plain" Block]{
		\begin{minipage}{0.3\linewidth}
			\centering
			\includegraphics[width=1in]{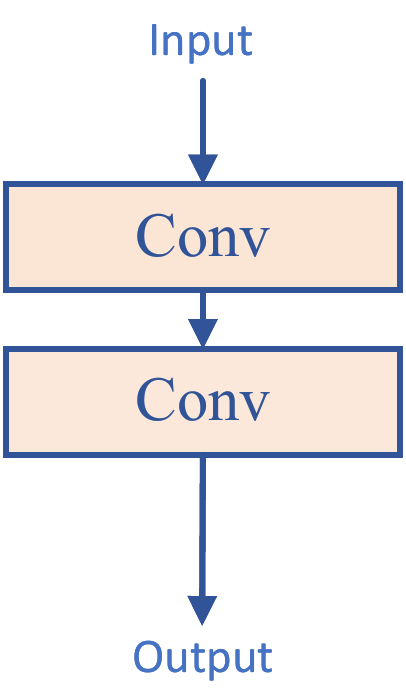}
		\end{minipage}%
	}%
	\subfigure[Basic Block]{
		\begin{minipage}{0.3\linewidth}
			\centering
			\includegraphics[width=1.8in]{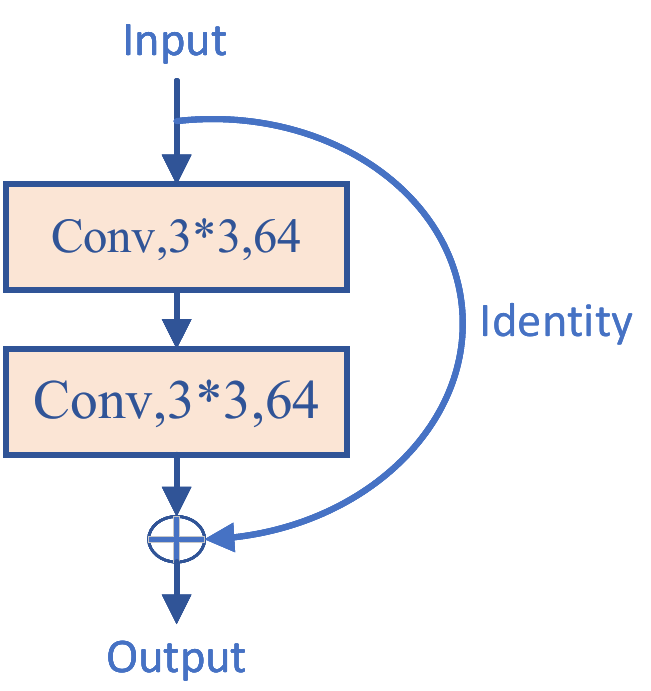}
		\end{minipage}%
	}%
	\subfigure[Bottle Block]{
		\begin{minipage}{0.3\linewidth}
			\centering
			\includegraphics[width=1.8in]{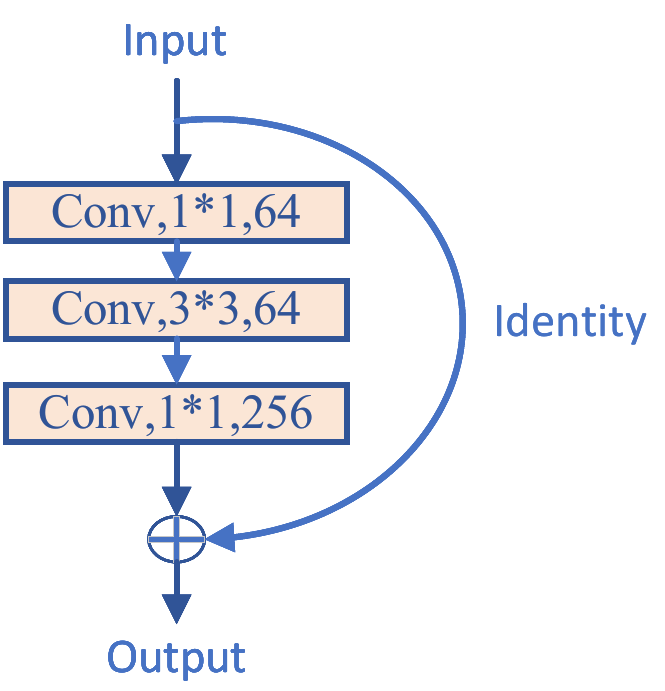}
		\end{minipage}
	}%
	\centering
	\caption{ Structure of three different base convolution blocks, where ``1*1'' and ``3*3" represent the size of the convolution kernel,
		``64" and ``256" represent the number of output  feature channels}
	\label{cnn}
\end{figure*}

\begin{figure*}[t]
	\centering
	\includegraphics[width=7in]{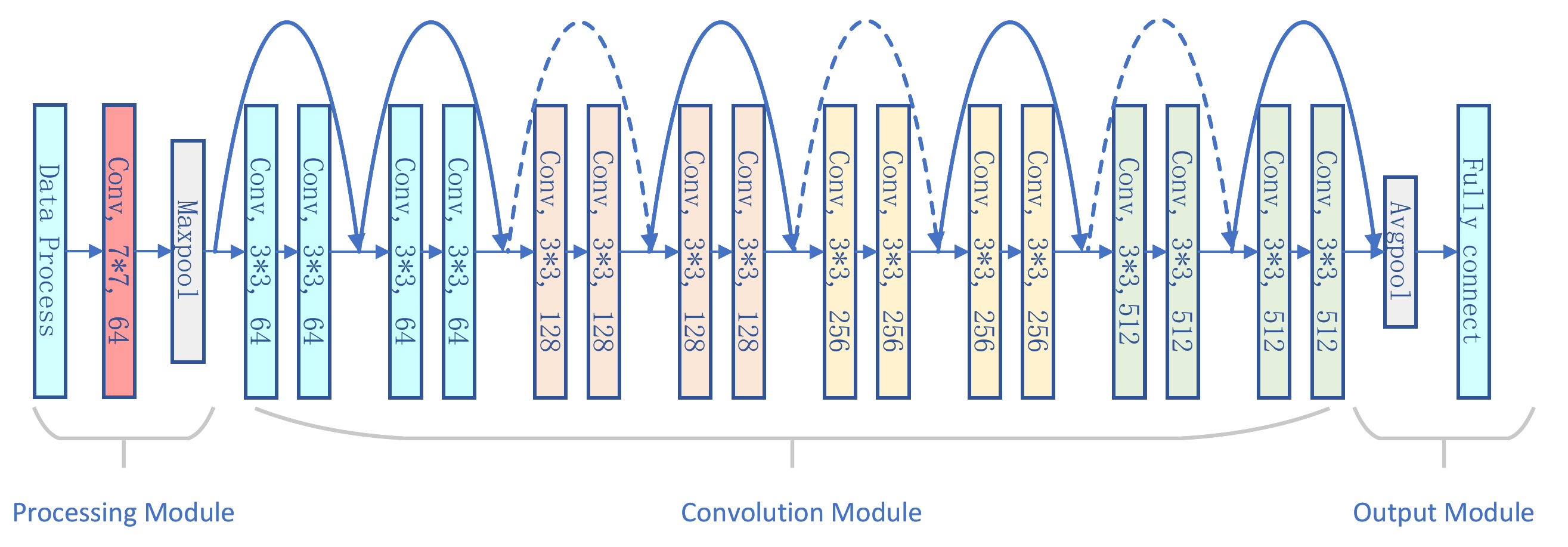}
	\caption{Structure of the residual neural network Resnet18. }
	\label{resnet18}\vspace{-0.9cm}
\end{figure*}
\vspace{-0.6cm}

\subsection{Deep Learning Model Design}

In this section, we describe the structure of the proposed neural network and its principles in details. Since the number of the codewords in the near-field codebook is finite, the estimation of the optimal codeword based on the received pilot signals is similar to the multi-classification problem. Consequently, convolutional neural networks (CNN) are employed for their powerful feature extraction capabilities and excellent classification performance. Moreover, based on the CNN, we further adopt the deep residual network (DRN), which is a variation of CNN and can solve the performance degradation problem when the number of CNN's layers is large. Specifically, in order to better extract the features in the input vector, the number of CNN's layers should be increased. However, as the number of CNN's layers increases to a certain extent, the performance of the neural network will degrade instead. This is due to the fact that the deeper the neural network is, the more pronounced the gradient disappearance is, which makes the parameters of the previous layers of the network not updated. To deal with this issue, residual networks are proposed, which consist of multiple residual blocks, to improve the performance of the network further.

$1)$ $\textit{Preprocessing Module}$: Before being fed into the neural network, the received signal vector is subject to data processing by the preprocessing module. Firstly, the received signal vector in complex form is divided into real part $\mathfrak{R}\left ( \textbf{y}^{\textrm{w}} \right )$ and an imaginary part  $\mathfrak{J}\left ( \textbf{y}^{\textrm{w}} \right )$, and both parts are transformed into a matrix with similar number of rows and columns.  Next, the first convolutional layer would upgrade the input from the two feature channels to the sixty-four feature channels by convolving. 

$2)$ $\textit{Convolution Module}$: In this module, massive convolution layers are deployed to extract features from the input received signal vector. For the residual networks, every two or three convolution layers would constitute the residual blocks first, and then the residual blocks are composed of the convolution module. As shown in Fig. \ref{cnn}, the convolution module of the residual network consists mainly of two basic residual blocks, namely Basic Block and Bottle Block, while the latter is designed for deeper networks.
In comparison to the regular convolution structure, the Basic Block and Bottle Block add identity mapping to the original base connection, which passes the current input directly to the next block. Such a structure, which is known as Short Connection, ensures that the gradient of the deeper network can be passed directly to the shallower network during backpropagation, thereby overcoming the problem of gradient disappearance to some extent. 

As an example, the structure of an 18-layer residual network, i.e. Resnet18, is shown in Fig. \ref{resnet18}. From Fig. \ref{resnet18}, it can be seen that the convolution module of Resnet18 consists of eight Basic Blocks, where the solid lines represent identity mappings of equal dimensions and the dashed lines represent identity mappings of different dimensions, which require additional convolution processing to convert to the same dimensions. The specific structural parameters of Resnet18 and Resnet50 are detailed in Table \ref{netwrok_table} in Section \ref{simulation}.

Note that each convolution layer is followed by a batch norm layer and a ReLU activation layer, whose purpose is to prevent overfitting and to increase the nonlinear mapping capability of the network, respectively. Furthermore, at the end of the convolution module, the average pooling layer is employed to transform the dimensionality of the output and map it to the fully connected layer.


$3)$ $\textit{ Output Module}$: Since the near-field deep learning-based beam training is similar to a multi-classification problem, a fully connected layer and a Softmax activation layer are deployed to perform the estimation of the optimal codeword. The function of the fully connected layer is to refine the features extracted by the convolution module and estimate the probability of each codeword being the optimal codeword, while the function of the Softmax activation layer is to normalize the output of the fully connected layer into probability values. Specifically, the function of the Softmax activation layer can be formulated as
\begin{equation}\label{softmax}
	\setlength\abovedisplayskip{3pt}
	\setlength\belowdisplayskip{3pt}
p_{i}=\frac{e^{a_{i}}}{\sum_{i=1}^{I}e^{a_{i}}},
\end{equation}
where $I$ denotes the output dimension of the fully connected layer; $p_{i}$ and $a_{i}$ represent the $i$-th output of the Softmax activation layer and the fully connected layer, respectively. Since the proposed scheme employs the dual neural network structure, i.e. x-axis network and y-axis network, to estimate the optimal near-field codeword separately, two probability distribution vectors would be obtained, which can be formulated as
\begin{equation}\label{net_out}
	\setlength\abovedisplayskip{3pt}
	\setlength\belowdisplayskip{3pt}
	\begin{gathered}
		\hat{\boldsymbol{P}}^{\textrm{x}}=\left [ \hat{p}_{1}^{\textrm{x}},\hat{p}_{2}^{\textrm{x}},\cdots ,\hat{p}_{S_{x}}^{\textrm{x}}\right ]^{T},\\
		\hat{\boldsymbol{P}}^{\textrm{y}}=\left [ \hat{p}_{1}^{\textrm{y}},\hat{p}_{2}^{\textrm{y}},\cdots ,\hat{p}_{S_{y}}^{\textrm{y}}\right ]^{T},
	\end{gathered}
\end{equation}
where $\hat{p}_{s_{x}}^{\textrm{x}}$ and $\hat{p}_{s_{y}}^{\textrm{y}}$ denote the probabilities that the x-axis index of the optimal near-field codeword is $s_{x}$ and the y-axis index is $s_{y}$, respectively. The greater the values of  $\hat{p}_{s_{x}}^{\textrm{x}}$ and $\hat{p}_{s_{y}}^{\textrm{y}}$, the higher the probability that the optimal near-field codeword is $\bar{\mathbf{c}}_{r}\left( x_{s_{x}},y_{s_{y}}\right )$.

Furthermore, the cross-entropy loss function will be used as an evaluation criterion, which is widely applied in multi-classification problems. Specifically, the cross-entropy loss function of the x-axis network and the y-axis network can be expressed as
\begin{equation}\label{loss_x_y}
	\setlength\abovedisplayskip{3pt}
	\setlength\belowdisplayskip{3pt}
	\begin{gathered}
		Loss^{\textrm{x}}=-\sum_{s_{x}=1}^{S_{x}}p_{s_{x}}^{\textrm{x}}\log_{10}\hat{p}_{s_{x}}^{\textrm{x}},\\
		Loss^{\textrm{y}}=-\sum_{s_{y}=1}^{S_{x}}p_{s_{x}}^{\textrm{y}}\log_{10}\hat{p}_{s_{y}}^{\textrm{y}},
	\end{gathered}
\end{equation}
where $p_{s_{x}}^{\textrm{x}}=1$ and ${p}_{s_{y}}^{\textrm{y}}=1$ if the actual optimal near-field codeword is  $\bar{\mathbf{c}}_{r}\left( x_{s_{x}},y_{s_{y}}\right )$. Otherwise, $p_{s_{x}}^{\textrm{x}}=0$ and ${p}_{s_{y}}^{\textrm{y}}=0$.

\vspace{-0.6cm}
\subsection{Far-field Beam based Beam Training }

\begin{figure*}[t]
	\centering
	\includegraphics[width=6in]{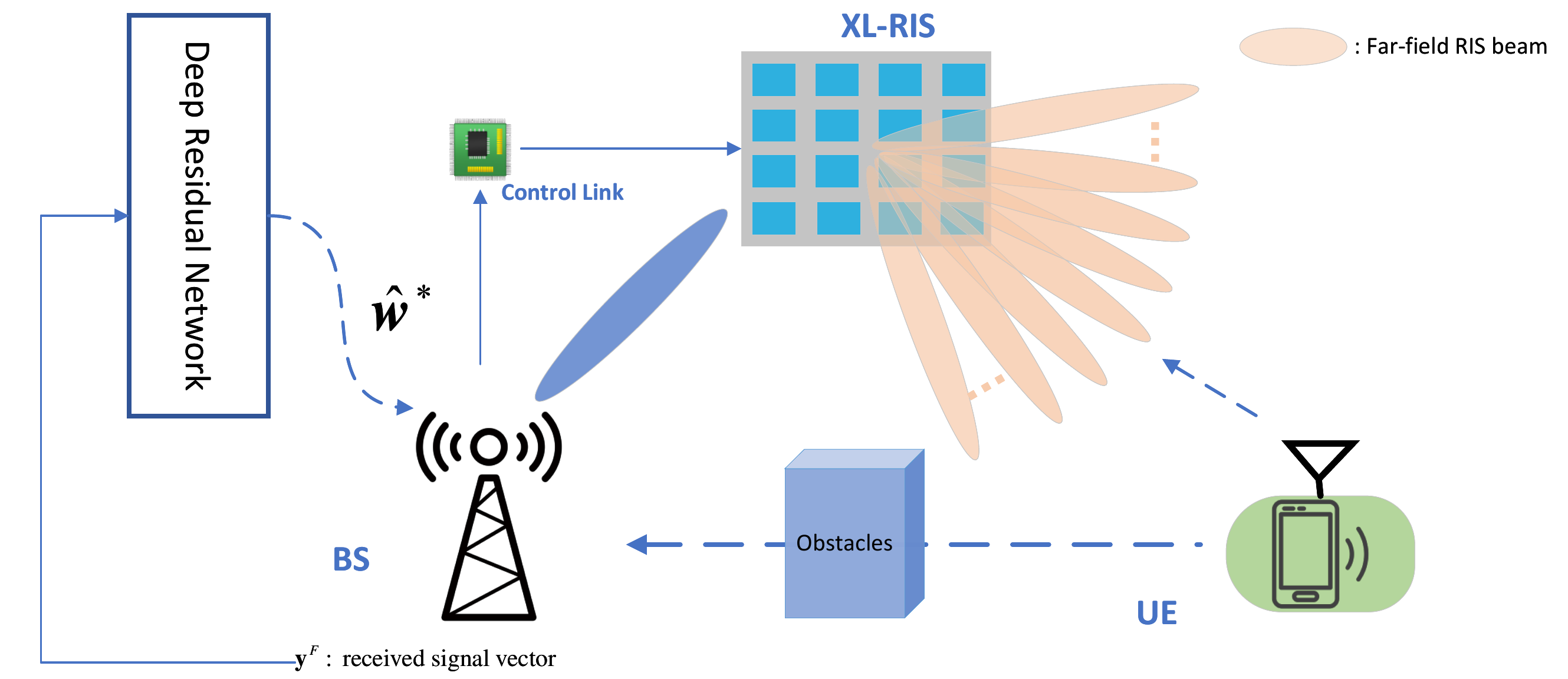}
	\caption{The framework of the proposed FBT scheme in which all far-field RIS codewords are tested . }
	\label{fbt}
	\vspace{-0.9cm}
\end{figure*}

Given the detailed design of the residual network structure in the previous section, two primary beam training schemes are developed, namely the FBT scheme and PNBT scheme, which differ in the choice of the codewords to be tested in the pilot transmission phase.
In this section, the detailed steps of the FBT scheme will be provided.

The framework of the proposed FBT scheme is shown in Fig. \ref{fbt}. For the FBT scheme, the user first sends the pilot signals to the BS via the XL-RIS, during which the XL-RLS selects codewords from the far-field RIS codebook to form the phase shifts at different time slots. By defining the receive signal  of the $n$-th far-field RIS codeword as $y_{n}^{F}$, the BS can obtain the receive signal vector $\textbf{y}^{F}=\left [ y_{1}^{F},y_{2}^{F}\cdots y_{N}^{F} \right ]^{T}$. According to (\ref{re_y_q}), the received signal $y_{n}^{F}$ can be expressed as
\begin{equation}\label{re_y_f}
	\setlength\abovedisplayskip{1pt}
	\setlength\belowdisplayskip{1pt}
		y_{n}^{F}=\beta\boldsymbol{f}_{n}^{T}\bar{\mathbf{c}}_{r}\left( x^{k}_{1},y^{k}_{1},z^{k}_{1}\right )x+n_{e,n},
\end{equation}
where $\boldsymbol{f}_{n}$ denotes the $n$-th codeword in the far-field RIS codebook $\mathcal{F}$. Note that since all far-field RIS codewords are tested, the total time slot length $Q$ is equal to the number of far-field RIS codewords $N$.

Then, the received signal vector $\textbf{y}^{F}$ are fed into the x-axis network and the y-axis network respectively and the probability distribution vectors  $\hat{\boldsymbol{P}}^{\textrm{x}}$ and $\hat{\boldsymbol{P}}^{\textrm{y}}$ can be obtained according to (\ref{net_out}). Based on $\hat{\boldsymbol{P}}^{\textrm{x}}$ and $\hat{\boldsymbol{P}}^{\textrm{y}}$, the x-axis index and y-axis index of the optimal near-field codeword can be expressed as
\begin{equation}\label{index_or1}
	\setlength\abovedisplayskip{3pt}
	\setlength\belowdisplayskip{3pt}
	\begin{gathered}
	\hat{s}_{x}^{\star }=\arg\max_{s_{x}=1,2,\cdots ,S_{x}}[\hat{\boldsymbol{P}}^{\textrm{x}}]_{s_{x}},
	\\
	\hat{s}_{y}^{\star }=\arg\max_{s_{y}=1,2,\cdots ,S_{y}}[\hat{\boldsymbol{P}}^{\textrm{y}}]_{s_{y}}.
	\end{gathered}
\end{equation}
Then  the index of the optimal near-field codeword in the codebook can be formulated as 
\begin{equation}\label{singal_w_set_1}
	\setlength\abovedisplayskip{3pt}
	\setlength\belowdisplayskip{3pt}
	\hat{s}^{\star }=\left ( \hat{s}_{y}^{\star}-1 \right )S_{x}+\hat{s}_{x}^{\star }.
\end{equation}
Finally, the optimal near-field RIS cordword can be obtained by
\vspace{-0.6cm}
\begin{eqnarray}\label{fbt_out}
	\hat{\boldsymbol{w}}^{\star }=[\mathcal{W}]_{\hat{s}^{\star }}.
\end{eqnarray}

\vspace{-0.7cm}

\begin{algorithm}[htb]
	\caption{ The Far-field Beam-based Training.}
	\label{alg:FBT}
	\begin{algorithmic}[1] 
		\REQUIRE ~~\\ 
		Well-trained x-axis and y-axis residual networks;\\
		\ENSURE ~~\\ 
		Optimal near-field codeword $\boldsymbol{w}^{\star }$;
		\STATE XL-RIS implements far-field beam test and the received signal vectors $\textbf{y}^{F}$ can be obtained at the BS according to (\ref{re_y_f});
		\STATE Obtain the probability distribution vector $\hat{\boldsymbol{P}}^{\textrm{x}}$ and $\hat{\boldsymbol{P}}^{\textrm{y}}$  by inputting $\textbf{y}^{F}$ to the x-axis and y-axis residual networks, respectively;
		\STATE The x-axis index and y-axis index of the optimal near-field codeword can be obtained via (\ref{index_or1});
		\STATE The optimal near-field codeword $\hat{\boldsymbol{w}}^{\ast }$ can be obtained via (\ref{fbt_out});
		\RETURN $\hat{\boldsymbol{w}}^{\star }$; 
	\end{algorithmic}

\end{algorithm}
\vspace{-0.9cm}
\subsection{Partial Near-field Beam-based Training }


$1)$ $\textit{ Partial Near-field Beam-based Training Scheme}$: 

Compared to the FBT scheme, the XL-RIS in the PNBT scheme selects the partial near-field codewords for test at step 2. Specifically, the XL-RIS selects the codewords from the near-field RIS codebook at equally spaced intervals according to the index to form the phase shifts at different pilot time slots. By defining $D$ as the sampling interval and $I=\left \lfloor \frac{S_{x}S_{y}}{D}\right \rfloor$ as the number of codewords picked, the set of selected near-field RIS codewords can be represented as
\begin{equation}\label{pnbt_test_beam}
	\setlength\abovedisplayskip{3pt}
	\setlength\belowdisplayskip{3pt}
\mathcal{D} = \left\{\boldsymbol{w}_{t_{i}}\big|t_{i}=i\times D,i=1,2,\cdots ,I\right\},
\end{equation}
where $\boldsymbol{w}_{t_{i}}$ denotes the $t_{i}$-th codeword in the near-field RIS codebook $\mathcal{W}^{N}$. 

Then, according to (\ref{re_y_q}), the received signal of the $t_{i}$-th near-field codeword can be represented as\vspace{-0.3cm}
\begin{equation}\label{re_y_n}
	\setlength\abovedisplayskip{3pt}
	\setlength\belowdisplayskip{3pt}
	y_{t_{i}}^{N}=\beta\boldsymbol{w}_{t_{i}}^{T}\bar{\mathbf{c}}_{r}\left( x^{k}_{1},y^{k}_{1},z^{k}_{1}\right )x+n_{e,t_{i}},
	\vspace{-0.3cm}
\end{equation}
and the received signal vector $\textbf{y}^{N}=\left [ y_{t_{1}}^{N},y_{t_{2}}^{N}\cdots y_{t_{I}}^{N} \right ]^{T}$ can be obtained at the BS.

Furthermore, the remaining steps for obtaining the optimal near-field codeword in the PNBT scheme are the same as the FBT scheme and will not be repeated here.

\vspace{-0.3cm}
\begin{algorithm}[htb]
	\caption{ The Partial Near-field Beam-based Training.}
	\label{alg:PNBT}
	\begin{algorithmic}[1] 
		\REQUIRE ~~\\ 
	    Well-trained x-axis and y-axis residual networks;\\
	\ENSURE ~~\\ 
	Optimal near-field codeword, $\boldsymbol{w}^{\star }$;
	\STATE XL-RIS implements partial near-field codewords test and the received signal vectors $\textbf{y}^{N}$ can be obtained at the BS according to (\ref{re_y_n});
	\STATE Obtain the probability distribution vector $\hat{\boldsymbol{P}}^{\textrm{x}}$ and $\hat{\boldsymbol{P}}^{\textrm{y}}$  by inputting $\textbf{y}^{N}$ to the x-axis and y-axis residual networks, respectively;
	\STATE The x-axis index and y-axis index of the optimal near-field codeword can be obtained via (\ref{index_or1});
	\STATE The optimal near-field codeword $\hat{\boldsymbol{w}}^{\ast }$ can be obtained via (\ref{fbt_out});
	\RETURN $\hat{\boldsymbol{w}}^{\star }$; 
	\end{algorithmic}
\end{algorithm}
\vspace{-0.7cm}
$2)$ $\textit{ The Improved PNBT Scheme}$:

\begin{figure*}[t]
	\centering
	\includegraphics[width=6.2in]{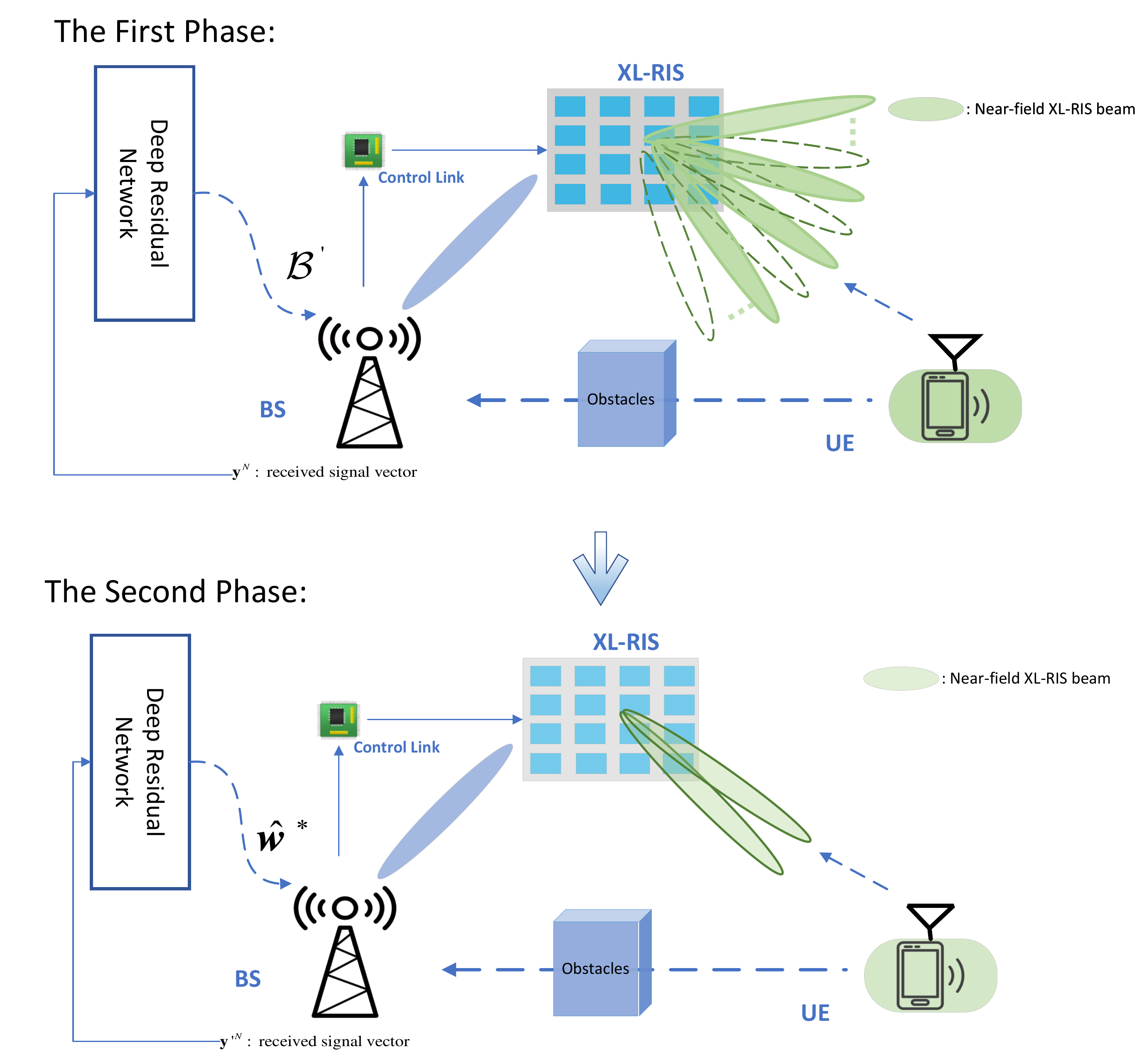}
	\caption{The framework of the improved PNBT scheme in which only partial near-field XL-RIS codewords are tested in the first phase and only codewords in set $\mathcal{B}^{'}$ are tested in the second phase. }
	\label{pnbt}\vspace{-0.9cm}
\end{figure*}

The PNBT scheme can significantly reduce the pilot overhead; however, the estimation performance will degrade when the sampling interval $D$ is large. In order to further improve the performance of the PNBT scheme, we propose an improved PNBT scheme in which the output of the neural network, i.e. $\hat{\boldsymbol{P}}^{\textrm{x}}$ and $\hat{\boldsymbol{P}}^{\textrm{y}}$, will be further exploited and additional codeword tests are required. The framework of the improved PNBT scheme is shown in Fig. \ref{pnbt}.

As shown in the diagram of Algorithm 3, the steps to obtain the probability distribution vectors $\hat{\boldsymbol{P}}^{\textrm{x}}$ and $\hat{\boldsymbol{P}}^{\textrm{y}}$ are the same for both the PNBT scheme and the improved PNBT scheme. Based on the $\hat{\boldsymbol{P}}^{\textrm{x}}$ and $\hat{\boldsymbol{P}}^{\textrm{y}}$, the improved PNBT scheme performs additional tests on some of the more possible near-field codewords rather than directly selecting the most possible codeword, which differs from the original scheme. Specifically, the BS can obtain the first $K$ maximum possible x-axis indices and $L$ maximum possible y-axis indices according to $\hat{\boldsymbol{P}}^{\textrm{x}}$ and $\hat{\boldsymbol{P}}^{\textrm{y}}$, which can be clearly represented in the set $\mathcal{L}_{\mathrm{x}}$ and $\mathcal{L}_{\mathrm{y}}$ as 
\vspace{-0.5cm}
\begin{align}\label{order}
	\left\{ \hat{p}_{\sigma _{1}}^{\textrm{x}},\hat{p}_{\sigma_{2}}^{\textrm{x}},\cdots ,\hat{p}_{\sigma _{S_{x}}}^{\textrm{x}}\right\}&=\left< \left\{\hat{p}_{1}^{\textrm{x}},\hat{p}_{2}^{\textrm{x}},\cdots ,\hat{p}_{S_{x}}^{\textrm{x}} \right\}\right>,
		\\
	\left\{ \hat{p}_{\gamma _{1}}^{\textrm{y}},\hat{p}_{\gamma_{2}}^{\textrm{y}},\cdots ,\hat{p}_{\gamma _{S_{y}}}^{\textrm{y}}\right\}&=\left< \left\{\hat{p}_{1}^{\textrm{y}},\hat{p}_{2}^{\textrm{y}},\cdots ,\hat{p}_{S_{y}}^{\textrm{y}} \right\}\right>,
	\\
	\mathcal{L}_{\mathrm{x}}&=\left \{ \sigma _{1},\sigma _{2},\cdots ,\sigma _{K} \right \},
	\\
	\mathcal{L}_{\mathrm{y}}&=\left \{ \gamma _{1},\gamma _{2},\cdots ,\gamma _{L} \right \},
\end{align}
where $\left \langle \cdot  \right \rangle$ denotes the  order operation, e.g., for $\mathcal{A}=\left\{a_{1}, a_{2}, \ldots, a_{n}\right\},\langle\mathcal{A}\rangle=\left\{a_{\sigma_{1}}, a_{\sigma_{2}}, \ldots, a_{\sigma_{n}}\right\}$ with $a_{\sigma_{1}} \geq a_{\sigma_{2}} \geq \ldots \geq a_{\sigma_{n}}$.  Based on $\mathcal{L}_{\mathrm{x}}$ and $	\mathcal{L}_{\mathrm{y}}$, the codewords that point to the intersection of these indices can be represented by the set $\mathcal{B}$ as
\vspace{-0.2cm}
\begin{equation}\label{set_B}
	\setlength\abovedisplayskip{3pt}
	\setlength\belowdisplayskip{3pt}
	\mathcal{B} = \left\{\boldsymbol{w}_{b_{j}}\big|{b}_{j}=(\gamma -1)S_{x}+\sigma ,\sigma \in 	\mathcal{L}_{\mathrm{x}},\gamma \in \mathcal{L}_{\mathrm{y}},\\j=1,2, \ldots, KL\right\}.
\end{equation}

In order to avoid duplicate test, the set $\mathcal{B}$ requires to exclude the near-field codewords from the first test, and eventually, the additional  codewords that need to be tested can be represented as 
\vspace{-0.5cm}
\begin{equation}\label{set_B_1}
	\setlength\abovedisplayskip{3pt}
	\setlength\belowdisplayskip{3pt}
	\mathcal{B}^{'}=\mathcal{B}-\mathcal{B}\cap \mathcal{D}.
	\vspace{-0.5cm}
\end{equation}
Then the BS delivers the set  $\mathcal{B}^{'}$ to the XL-RIS via the direct control link, and the user needs to transmit a few pilot signals again for additional near-field codewords test, whose procedures are the same as that for the first test. By implementing additional near-field codewords test, the received signal vector corresponding to the near-field codewords with the maximum probability can be obtained as
\vspace{-0.6cm}
\begin{align}\label{ad_re_signal}
y_{b_{j}}^{N}&=\beta\boldsymbol{w}_{b_{j}}^{T}\bar{\mathbf{c}}_{r}\left( x^{k}_{1},y^{k}_{1},z^{k}_{1}\right )x+n_{e,n},\\
\mathbf{y'}^{\textrm{N}}&=\left [y_{b_{1}}^{N},y_{b_{2}}^{N} ,\cdots , y_{b_{KL}}^{N}\right ]^{T}.
\vspace{-0.cm}
\end{align}
Finally, based on the measurement of the received signal vector, the index of the optimal near-field codeword in the codebook $\mathcal{W}$  can be represented as
\begin{equation}\label{index_n}
	\setlength\abovedisplayskip{3pt}
	\setlength\belowdisplayskip{3pt}
s^{\star }=\mathop{\arg\max}\limits_{b_{j}\in \mathcal{B}}\left | y_{b_{j}}^{N} \right |,
\end{equation}
and the optimal near-field codeword can be obtained as
\begin{equation}\label{op_n}
	\setlength\abovedisplayskip{1pt}
	\setlength\belowdisplayskip{3pt}
\hat{\boldsymbol{w}}^{\star }=[\mathcal{W}]_{s^{\star }}.
\end{equation}

\vspace{-0.5cm}
\begin{algorithm}[htb]
	\caption{ The Improved PNBT.}
	\label{alg:Framwork}
	\begin{algorithmic}[1] 
		\REQUIRE ~~\\ 
		Well-trained x-axis and y-axis residual networks;\\
		The number of additional codewords to be tested $K,L$;\\
		\ENSURE ~~\\ 
		Optimal near-field codeword, $\boldsymbol{w}^{\star}$;
		\STATE XL-RIS implements partial near-field codewords test and the received signal vectors $\textbf{y}^{N}$ can be obtained at the BS according to (\ref{re_y_n});
		\STATE Obtain the probability distribution vector $\hat{\boldsymbol{P}}^{\textrm{x}}$ and $\hat{\boldsymbol{P}}^{\textrm{y}}$  by inputting $\textbf{y}^{N}$ to the x-axis and y-axis residual network, respectively;
		\STATE The set of near-field codewords that require additional test 	$\mathcal{B}^{'}$ can be obtained by (\ref{set_B}) and (\ref{set_B_1});
		\STATE The XL-RIS performs additional beam test and $\mathbf{y'}^{\textrm{N}}$ can be obtained by (\ref{ad_re_signal});
		\STATE The index of the optimal near-field codeword $s^{\star }$ can be obtained by (\ref{index_n});
		\STATE  The optimal near-field codeword $\boldsymbol{w}^{\star }$ can be obtained by (\ref{op_n})
		\RETURN $\hat{\boldsymbol{w}}^{\star}$; 
	\end{algorithmic}
\end{algorithm}
\vspace{-0.7cm}


\section{Simulation Results}\label{simulation}
In order to evaluate the performances of the two DRN-based near-field beam training schemes, numerical simulations are implemented, and the simulation results are presented in this section. Firstly, the parameters of the simulation system are presented, which include the parameters of the channel, the codebook and the neural networks. Then, the simulation results of the proposed scheme are given and compared with the existing schemes. In addition, three performance metrics are proposed to assess the performances of beam training schemes.

\vspace{-0.6cm}
\subsection{System Setup}

In our simulations, XL-RIS-assisted mmWave communication systems are considered, where the number of antennas of the BS and the number of elements of the XL-RIS are set as $M$=512 and $N$=512, respectively. The number of channel paths from the BS to the XL-RIS and from the XL-RIS to the user are set as $L_{G}$=3 and $L_{k}$=3 respectively, where the channel gains of the strongest path are set as $\alpha ^{G}_{1}, \alpha ^{k}_{1}\sim \mathcal{C} \mathcal{N}(0,1)$ and the gains of the remaining channels are set as $\alpha ^{G}_{l_{1}}, \alpha ^{k}_{l_{2}}\sim \mathcal{C} \mathcal{N}(0,0.001)$ for $l_{1}, l_{2}=2,3$. Furthermore, the range of the user's plane coordinate $\left( x_{1},y_{1}\right )$ is given by $x_{1}\in (-50\ m,50\ m)$ and $y_{1}\in(-30\ m,30\ m)$, which are within the limits of the near-field domain. The sampling intervals of the near-field RIS codebook in the x-axis and y-axis are set as $\Delta _{x}=1m$ and $\Delta _{y}=1m $, which results in the numbers of sampling points being $S_{x}=100$ and $S_{y}=60$, respectively. Consequently, the number of candidate codewords in the near-field XL-RIS codebook reaches $S_{x}S_{y}=6,000$, which is far more than the number of far-field codewords $N=512$. Finally, the selection interval for the PNBT scheme is fixed to $D=20$, which means that only 1 in 20 codewords in the PTBT scheme is tested. Based on the above channel and codebook parameters, we generate 20,000 channel samples and corresponding labels, of which 90 percentage are used for training the neural network and 10 percentage for evaluating the performance.

For the proposed neural network, 18-layer and 50-layer residual networks, i.e. Resnet18 and Resnet50, are adopted, and the specific structural parameters are shown in Table \ref{netwrok_table}. In order to accelerate the convergence of the neural networks, the learning rate decay strategy is employed, where the initial learning rate is set to 0.005, and the learning rate decays by half when the estimation accuracy does not improve within two training epochs. Moreover, the mini-batch strategy is also adopted, where the neural network is trained for 120 epochs with 2000 batches in each epoch. Fig. \ref{loss} shows the variation of the loss function with training epochs during the training phase. The loss gradually decreases and flattens out at approximately epoch 40. Then, the model parameters of the well-trained x-axis network and y-axis network are saved and used for subsequent performance tests.

\begin{table}[t]
	\centering \label{table1} \caption{\text { DEFAULT SYSTEM PARAMETERS }}
	\begin{tabular}{|c|c|}
		\hline
		\textbf{Parameters}                                & \textbf{Value}                               \\ \hline
		Carrier frequency                                  & 30 GHz                                       \\ \hline
		Number of  antennas in BS                           & 512                                          \\ \hline
		Number of elements in XL-RIS                       & 512                                      \\ \hline
		Number of RIS far-field beams                      & 512                                          \\ \hline
		Number of RIS near-field beams                      & 6000                                        \\ \hline
		
		Sampling interval $\Delta _{x}, \Delta _{y}$                                      & 1m                                            \\ \hline
		Selection interval $D$                             &20                                \\ \hline
		The distribution of $x_{1}$                        & $\mathit{u}(-50 \mathrm{~m}, 50 \mathrm{~m})$ \\ \hline
		The distribution of $y_{1}$                        & $\mathit{u}(-30 \mathrm{~m}, 30 \mathrm{~m})$ \\ \hline
		Coordinates of the BS                     & $(20\mathrm{~m},20\mathrm{~m},0\mathrm{~m})$ \\ \hline
		Antenna spacing                    & $\frac{\lambda }{2}$ \\ \hline
	\end{tabular}
\vspace{-0.5cm}
\end{table}

\begin{table}[t]
	\centering 
	\caption{\text { NETWORK PARAMETERS }} 
	\label{netwrok_table} 
	\resizebox{0.6\columnwidth}{!}{
\begin{tabular}{|c|cc|}
	\hline
	Module name                         & \multicolumn{1}{c|}{18-layer}                                                                                       & 50-layer                                                                                                                 \\ \hline
	\multirow{2}{*}{Processing Module}  & \multicolumn{2}{c|}{Conv, 7*7, 64}                                                                                                                                                                                                             \\ \cline{2-3} 
	& \multicolumn{2}{c|}{Max pool}                                                                                                                                                                                                                  \\ \hline
	\multirow{4}{*}{Convolution Module} & \multicolumn{1}{c|}{$\left[\begin{array}{l}Conv, 3 \times 3,64 \\Conv, 3 \times 3,64\end{array}\right] \times 2$}   & $\left[\begin{array}{c}Conv, 1 \times 1,64 \\ Conv, 3 \times 3,64 \\ Conv, 1 \times 1,256\end{array}\right] \times 3$    \\ \cline{2-3} 
	& \multicolumn{1}{c|}{$\left[\begin{array}{l}Conv, 3 \times 3,128 \\Conv, 3 \times 3,128\end{array}\right] \times 2$} & $\left[\begin{array}{c}Conv, 1 \times 1,128 \\ Conv, 3 \times 3,128 \\ Conv, 1 \times 1,512\end{array}\right] \times 4$  \\ \cline{2-3} 
	& \multicolumn{1}{c|}{$\left[\begin{array}{l}Conv, 3 \times 3,256 \\Conv, 3 \times 3,256\end{array}\right] \times 2$} & $\left[\begin{array}{c}Conv, 1 \times 1,256 \\ Conv, 3 \times 3,256 \\ Conv, 1 \times 1,1024\end{array}\right] \times 6$ \\ \cline{2-3} 
	& \multicolumn{1}{c|}{$\left[\begin{array}{l}Conv, 3 \times 3,256 \\Conv, 3 \times 3,256\end{array}\right] \times 2$} & $\left[\begin{array}{c}Conv, 1 \times 1,512 \\ Conv, 3 \times 3,512 \\ Conv, 1 \times 1,2048\end{array}\right] \times 3$ \\ \hline
	\multirow{2}{*}{Output Module}      & \multicolumn{2}{c|}{Average pool}                                                                                                                                                                                                              \\ \cline{2-3} 
	& \multicolumn{2}{c|}{Fully connect (softmax)}                                                                                                                                                                                                   \\ \hline
\end{tabular}}
\vspace{-0.9cm}
\end{table}

\vspace{-0.6cm}
\subsection{Metrics and Baselines}

In order to better evaluate and compare the performances of different near-field beam training schemes, three performance metrics are adopted.

1) Achievable rate $E$ is given by
\begin{equation}\label{criteria_1}
	\setlength\abovedisplayskip{3pt}
	\setlength\belowdisplayskip{3pt}
	E=  \log _{2}\left(1+\frac{\left|\hat{\boldsymbol{w}}^{\star T} \bar{\mathbf{c}}_{r}\left( x_{1},y_{1},z_{1}\right ) \right|^{2}}{\sigma^{2}}\right),
\end{equation}
where $\hat{\boldsymbol{w}}^{\star}$ denotes the optimal near-field codeword estimated by the beam training scheme.

2) Normalizd RIS beam gain $G$ is given by 
\begin{equation}\label{criteria_2}
	\setlength\abovedisplayskip{3pt}
	\setlength\belowdisplayskip{3pt}
		G=\frac{\left|\hat{\boldsymbol{w}}^{\star T} \bar{\mathbf{c}}_{r}\left( x_{1},y_{1},z_{1}\right ) \right|^{2}}{\left|\boldsymbol{w}^{\star T} \bar{\mathbf{c}}_{r}\left( x_{1},y_{1},z_{1}\right ) \right|^{2}},
\end{equation}
where ${\boldsymbol{w}}^{\star}$ denotes the actual optimal near-field codeword. The normalized RIS beam gain can better reflect the performance enhancement achieved by the deep learning compared to the estimation accuracy since even beams that are incorrectly estimated by the neural networks are often aligned in the vicinity of the target, which can still yield significant beam gain and should be counted as the contribution of the deep learning \cite{make1}.

3) Effective achieveable rate  $\bar{E}$ is given by
\begin{equation}\label{criteria_3}
	\setlength\abovedisplayskip{3pt}
	\setlength\belowdisplayskip{3pt}
	\bar{E}=  \left ( 1- \frac{T_{\mathrm{tra}}}{T_{\text {tot }}} \right )\log _{2}\left(1+\frac{\left|\hat{\boldsymbol{w}}^{\ast T} \bar{\mathbf{c}}_{r}\left( x_{1},y_{1},z_{1}\right ) \right|^{2}}{\sigma^{2}}\right),
\end{equation}
where $T_{\mathrm{tot}}$ represents the total number of time slots in one channel coherence interval and $T_{\text {tra }}$ represents the number of time slots for beam training in each channel coherence time interval. The effective achievable rate provides a good indication of the comprehensive performance of the beam training schemes. Only the scheme with a high achievable rate and low pilot overhead can achieve a higher effective achievable rate.

To further measure the performance of the proposed schemes, the following four baselines are adopted:

1) Exhaustive Beam Search: The exhaustive beam search is the most straightforward approach and is known as the sweeping scheme, where the XL-RIS tests all near-field codewords, and the BS selects the codeword corresponding to the received signal with the highest energy.

2) Hierarchical Beam Search: The authors of \cite{wei_codebook} proposed a hierarchical search scheme for the XL-RIS, and a multi-layer codebook was employed, where the range of the higher-level codewords can cover several lower-level codewords. In the hierarchical search scheme, the optimal higher-level codeword is obtained by the exhaustive test, and then the lower-level codewords covered by that codeword are exhaustively tested to find the optimal lower-level codeword.

3) ML-based beam training in \cite{qi}: \cite{qi} proposed a beam training scheme for the far-field mmWave systems, where the fully connected (FC) neural network was employed. We adopt the same network structure as well as the beam training strategy and apply them to the near-field XL-RIS-assisted wireless communication system.

4) ML-based beam training in \cite{make1}: A CNN based far-field beam training scheme was proposed in \cite{make1}. Similarly, the same network structure and a close beam training strategy will be applied to the near-field for the XL-RIS, after which its performance will be compared with that of the proposed schemes in this paper.

\vspace{-0.6cm}
\subsection{Simulation Result}

$1)$ $\textit{How is the performance of the proposed solution ?}$

\begin{figure}
	\begin{minipage}[t]{0.49\linewidth}
		\centering
		\includegraphics[width=3.5in]{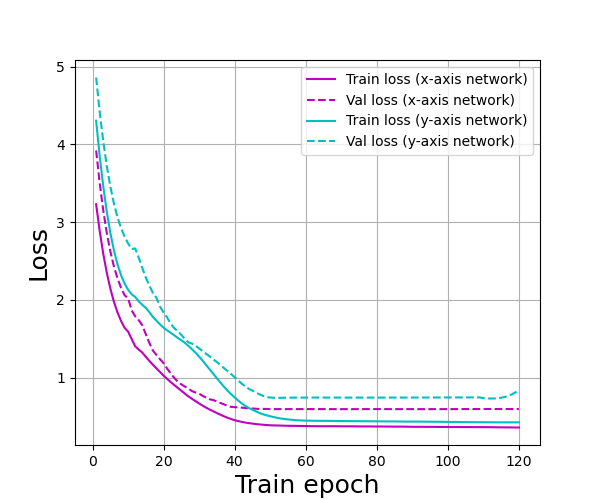}
		\vspace{-0.1cm}
		\caption{The loss function of training epoch.}
		\label{loss}\vspace{-0.5cm}
	\end{minipage}%
	\hfill
	\begin{minipage}[t]{0.49\linewidth}
		\centering
		\includegraphics[width=3.5in]{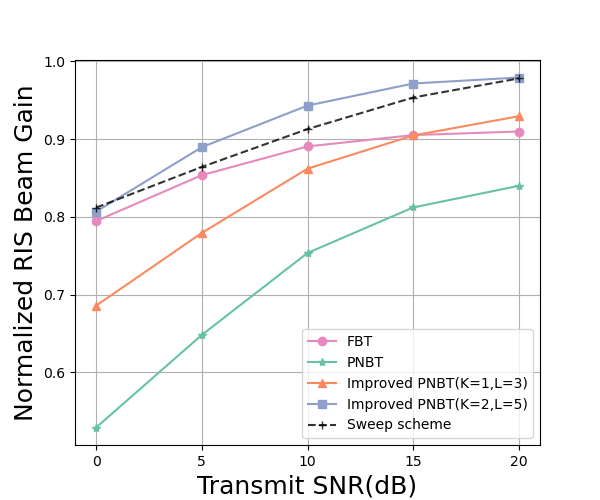}
		\vspace{-0.1cm}
		\caption{Normalized SNR for the proposed schemes at different transmit SNRs. }
		\label{norm}\vspace{-0.5cm}
	\end{minipage}%
	\hfill
	\vspace{-0.7cm}
\end{figure}

IN Fig. \ref{norm}, the performance of the proposed scheme in terms of the normalized RIS beam gain is compared at different transmission SNRs. From Fig. \ref{norm}, it can be seen that at higher transmit SNRs, e.g. SNR $>$ 5dB, the improved PNBT scheme with $K$=2 and $L$=5 can achieve more than 90\% of the perfect performance and even outperforms the sweeping scheme, which demonstrates the powerful non-linear relationship learning capability of the residual network. Furthermore, the FBT scheme can also have a similar beam gain performance to the sweeping scheme at different transmission SNRs. It can also be observed that the FBT scheme outperforms the PNBT scheme due to the fact that the FBT scheme tests more codewords, and the neural network can extract more information about the channel. However, with additional beam tests being implemented, the improved PNBT scheme gradually performs better than the FBT scheme because the knowledge of the probability distribution vector is fully exploited. 

$2)$ $\textit{How does the proposed scheme compare with the existing schemes?}$

\begin{figure}
	\begin{minipage}[t]{0.49\linewidth}
		\centering
		\includegraphics[width=3.5in]{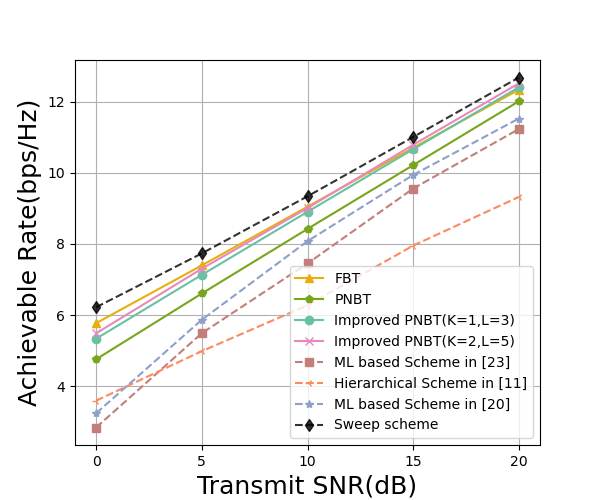}
		\vspace{-0.1cm}
		\caption{Achievable rate for the different beam training schemes. }
		\label{rate}\vspace{-0.5cm}
	\end{minipage}%
	\hfill
	\begin{minipage}[t]{0.49\linewidth}
		\centering
		\includegraphics[width=3.5in]{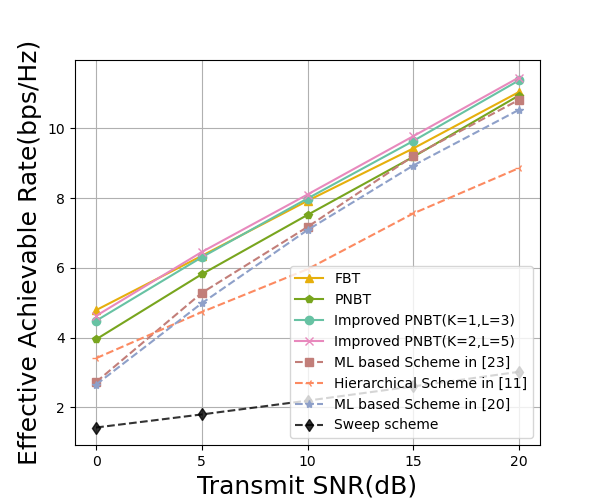}
		\vspace{-0.1cm}
		\caption{Effective  achievable rate for the different beam training schemes. }
		\label{e_rate}\vspace{-0.5cm}
	\end{minipage}%
	\hfill
	\vspace{-0.7cm}
\end{figure}

Fig. \ref{rate} compares the performances of the proposed schemes in terms of the achievable rate, where the existing beam training scheme is also included for comparison. As shown in Fig. \ref{rate}, both the proposed near-field beam training schemes exhibit similar performance to the sweeping scheme in terms of the achievable rate at all transmit SNRs; however, the pilot overhead is significantly less than that of the sweeping scheme. Furthermore, the proposed two near-field beam training schemes provide a significant performance improvement over the near-field hierarchical beam scheme because the hierarchical beam training is highly sensitive to noise and lacks the ability to handle it like the neural network. At the same time, the proposed schemes also outperform two ML-based beam training schemes, which are mainly suitable for the far-field domain and have difficulty in addressing the large amount of multi-dimensional near-field codewords.

Next, the performances of the different schemes in terms of achievable rate and pilot overhead, i.e. the effective achievable rate, are presented in Fig. \ref{e_rate}. As shown in Fig. \ref{e_rate}, although the sweeping scheme offers a high achievable rate performance, it suffers from a high pilot overhead, which results in a dramatic reduction in data communication time, as confirmed by its low effective achievable rate. For the proposed near-field beam training scheme, the FBT and PNBT schemes reduce the pilot overhead by 91\% and 95\%, compared to the sweeping scheme, where they only test 512 and 300 codewords, respectively. Moreover, the improved PNBT scheme with $K$=2 and $L$=5 only needs to test $I+KL=$310 codewords, thus reducing the pilot overhead by approximately 95\%. Consequently, the proposed near-field beam training scheme achieves excellent performance in terms of the effective achievable rates.

$3)$ $\textit{How about the impact of the number of codewords for additional test?}$

\begin{figure}
	\begin{minipage}[t]{0.48\linewidth}
		\centering
		\includegraphics[width=3.45in]{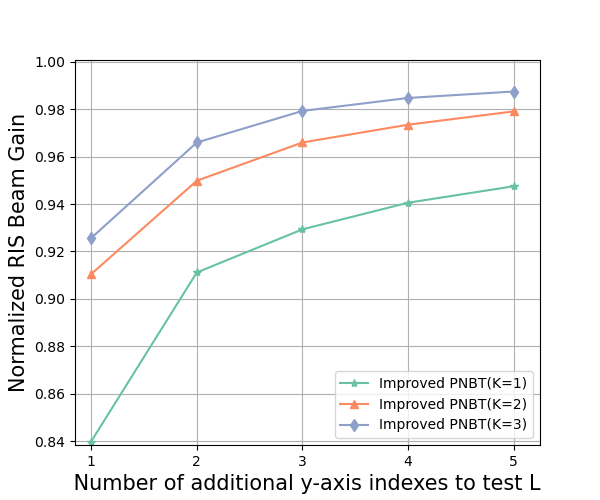}
		\vspace{-0.1cm}
		\caption{Normalizd SNR of the improved PNBT at different $K$,$L$.  }
		\label{kl}\vspace{-0.5cm}
	\end{minipage}%
	\hfill
	\begin{minipage}[t]{0.49\linewidth}
		\centering
		\includegraphics[width=3.5in]{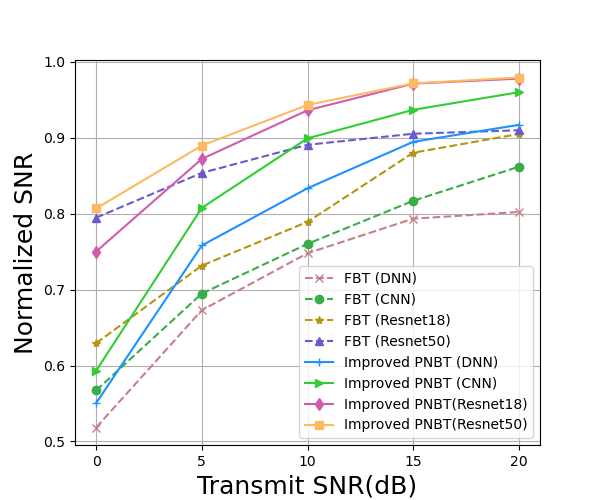}
		\vspace{-0.1cm}
		\caption{Normalizd SNR of the proposed schemes with different neural network structures. }
		\label{net}\vspace{-0.5cm}
	\end{minipage}%
	\hfill
	\vspace{-0.7cm}
\end{figure}

Fig. \ref{kl} illustrates the impact of different values of $K$ and $L$ on the normalized RIS beam gain of the improved PNBT scheme. It can be seen that the performance of the improved PNBT scheme improves as $K$ and $L$ increase because the knowledge of the probability distribution vectors is increasingly exploited. Additionally, the performance of the improved PNBT scheme tends to converge and approaches the perfect performance when $K>3$ and $L>3$, which means that the probability distribution vector has been fully exploited.

$4)$ $\textit{How is the performance of different neural network structures?}$

In Fig. \ref{net}, we focus on the impact of different neural network structures on the performance of the proposed near-field beam training scheme. Specifically, we employ different neural network structures with the same beam training strategy and compare their performances. It can be seen that the residual neural network-based near-field beam training schemes achieve more significant  RIS beam gain than the conventional CNN, which further demonstrates the superiority of residual networks. Additionally, with the increase of the residual network from 18 to 50 layers, the performance of the residual nerual network-based near-field beam training scheme does not degrade like the traditional convolutional network but further improves, which also proves the superior performance of the residual structure in solving the gradient disappearance problem.

$5)$ $\textit{Why we set the sampling interval D to 20?}$

\begin{figure}
	\begin{minipage}[t]{0.49\linewidth}
		\centering
		\includegraphics[width=3.5in]{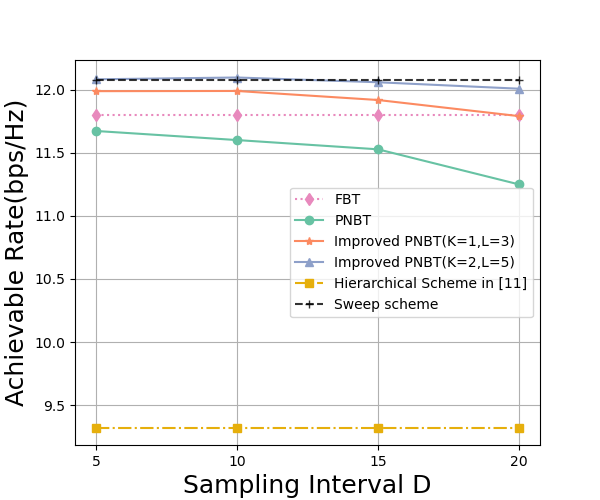}
		\vspace{-0.1cm}
		\caption{Achievable rate for the proposed schemes with different intervals $D$. }
		\label{rate_sample}\vspace{-0.5cm}
	\end{minipage}%
	\hfill
	\begin{minipage}[t]{0.49\linewidth}
		\centering
		\includegraphics[width=3.5in]{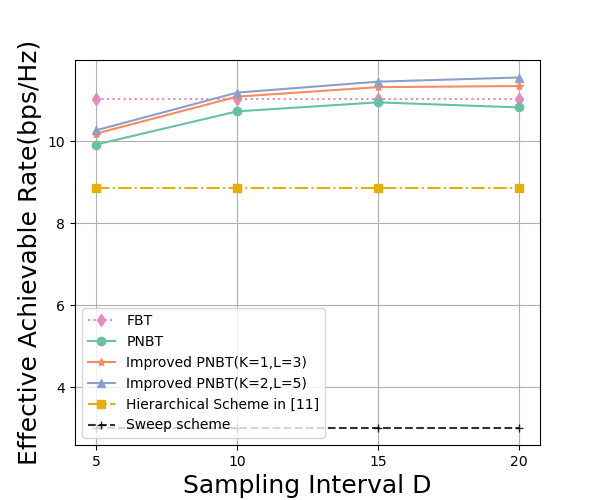}
		\vspace{-0.1cm}
		\caption{Effective  achievable rate for the proposed schemes with different intervals $D$.  }
		\label{e_rate_sample}\vspace{-0.5cm}
	\end{minipage}%
	\hfill
	\vspace{-0.7cm}
\end{figure}

Fig. \ref{rate_sample} and Fig. \ref{e_rate_sample} compare the performances of the proposed PNBT scheme with the improved PNBT scheme in terms of achievable rate and the effective achievable rate at different sampling intervals, respectively. From Fig. \ref{rate_sample}, it can be seen that the achievable rate of the PNBT scheme decreases correspondingly as the sampling interval increases, which is due to the fact that an increase in the sampling interval reduces the input to the neural network and thus the information acquired is reduced. Furthermore, the improved PNBT scheme receives less impact when the sampling interval is increased, especially when $K$=2 and $L$=5, the achievable rate of the improved PNBT scheme remains more or less the same and close to that of the sweeping scheme. In terms of the effective achievable rate, both the PNBT scheme and the improved PNBT scheme improve as the sampling interval increases, which is because the amount of pilot overhead is significantly reduced. Moreover, when the sampling interval reaches 20, the effective reachable rate of both the PNBT scheme and the improved PNBT scheme reaches their maximum value and start to converge, which means that the pilot overhead and achievable rate are in excellent balance. Therefore, we set the sampling interval to 20 in order to achieve a good performance of the proposed schemes.

$6)$ $\textit{How is the pilot overhead of the proposed scheme?}$

\begin{figure}[t]
	\centering
	\includegraphics[width=3.5in]{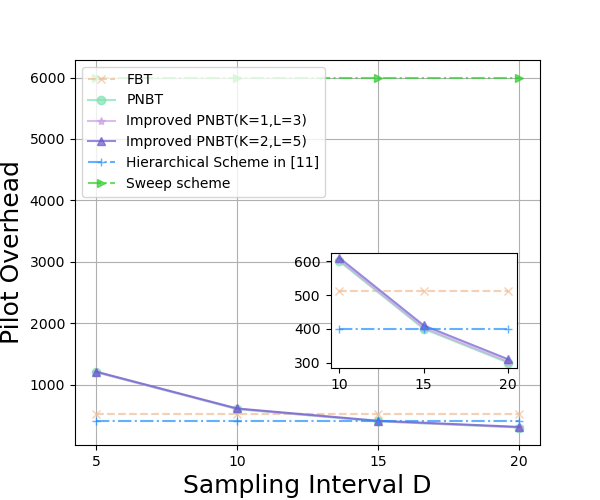}
	\vspace{-0.2cm}
	\caption{Normalizd SNR of the proposed schemes with different neural network structures. }
	\label{pilot_overhead}\vspace{-0.9cm}
\end{figure}

Fig. \ref{pilot_overhead} compares the pilot overhead of the proposed schemes with the existing beam training schemes. It can be seen that the proposed schemes significantly reduce the pilot overhead in the near-field XL-RIS-assisted wireless communication systems. When the sampling interval is 20, the proposed schemes are capable of reducing the pilot overhead by 91\% or even 95\%, and less than that of the hierarchical search scheme.

\vspace{-0.6cm}
\section{Conclusion}\label{conclusion}
In this paper, we proposed two deep learning-based near-field beam training schemes in XL-RIS-assisted wireless communication systems to reduce the pilot overhead, where the deep residual networks are employed to determine the optimal near-field RIS codeword. Our proposed scheme mainly utilizes neural networks to extract the implicit information between the received signals of different codewords, thus reducing the reliance on the information obtained from the codeword test. Specifically, we first proposed a FBT scheme in which the received signals of all far-field RIS codewords are fed into the neural network to estimate the optimal near-field RIS codeword. In order to further reduce the pilot overhead, the PNBT scheme was proposed, where only the received signals corresponding to the partial near-field RIS codewords were served as input to the neural network. Furthermore, we further proposed the  improved PNBT scheme to enhance the performance of the beam training by fully exploring the neural network's output. Note that in both schemes a dual neural network structure was proposed, i.e., an x-axis network and a y-axis network to estimate the index of the two dimensions of the optimal codeword. Simulation results showed that our proposed scheme can approach the performance of the sweeping scheme in terms of the achievable rate, however, the pilot overhead was significantly less than that of the sweeping scheme. Furthermore, the proposed schemes are more compatible with the XL-RIS-assisted wireless communication systems than the existing beam-training schemes, where better performance was obtained and the pilot overhead was reduced.


\bibliographystyle{IEEEtran}
\vspace{-0.6cm}
\bibliography{myre}

\begin{thebibliography}{10}
\providecommand{\url}[1]{#1}
\csname url@samestyle\endcsname
\providecommand{\newblock}{\relax}
\providecommand{\bibinfo}[2]{#2}
\providecommand{\BIBentrySTDinterwordspacing}{\spaceskip=0pt\relax}
\providecommand{\BIBentryALTinterwordstretchfactor}{4}
\providecommand{\BIBentryALTinterwordspacing}{\spaceskip=\fontdimen2\font plus
\BIBentryALTinterwordstretchfactor\fontdimen3\font minus
  \fontdimen4\font\relax}
\providecommand{\BIBforeignlanguage}[2]{{%
\expandafter\ifx\csname l@#1\endcsname\relax
\typeout{** WARNING: IEEEtran.bst: No hyphenation pattern has been}%
\typeout{** loaded for the language `#1'. Using the pattern for}%
\typeout{** the default language instead.}%
\else
\language=\csname l@#1\endcsname
\fi
#2}}
\providecommand{\BIBdecl}{\relax}
\BIBdecl

\bibitem{RIS_1}
C.~Pan, H.~Ren, K.~Wang, J.~F. Kolb, M.~Elkashlan, M.~Chen, M.~Di~Renzo,
  Y.~Hao, J.~Wang, A.~L. Swindlehurst, X.~You, and L.~Hanzo, ``Reconfigurable
  intelligent surfaces for {6G} systems: Principles, applications, and research
  directions,'' \emph{IEEE Commun. Mag.}, vol.~59, no.~6, pp. 14--20, 2021.

\bibitem{RIS_2}
C.~Pan, H.~Ren, K.~Wang, M.~Elkashlan, A.~Nallanathan, J.~Wang, and L.~Hanzo,
  ``Intelligent reflecting surface aided {MIMO} broadcasting for simultaneous
  wireless information and power transfer,'' \emph{IEEE J. Sel. Areas Commun.},
  vol.~38, no.~8, pp. 1719--1734, 2020.

\bibitem{RIS_3}
C.~Pan, H.~Ren, K.~Wang, W.~Xu, M.~Elkashlan, A.~Nallanathan, and L.~Hanzo,
  ``Multicell {MIMO} communications relying on intelligent reflecting
  surfaces,'' \emph{IEEE Trans. Wireless Commun.}, vol.~19, no.~8, pp.
  5218--5233, 2020.

\bibitem{RIS_4}
S.~Shen, B.~Clerckx, and R.~Murch, ``Modeling and architecture design of
  reconfigurable intelligent surfaces using scattering parameter network
  analysis,'' \emph{IEEE Trans. Wireless Commun.}, vol.~21, no.~2, pp.
  1229--1243, 2022.

\bibitem{RIS_material}
T.~J. Cui, M.~Q. Qi, X.~Wan, J.~Zhao, and Q.~Cheng, ``Coding metamaterials,
  digital metamaterials and programmable metamaterials,'' \emph{Light: science
  \& applications}, vol.~3, no.~10, pp. e218--e218, 2014.

\bibitem{LS_1}
D.~Mishra and H.~Johansson, ``Channel estimation and low-complexity beamforming
  design for passive intelligent surface assisted {MISO} wireless energy
  transfer,'' in \emph{Proc. IEEE Int. Conf. Acoust.,Speech Signal Process.
  (ICASSP)}, 2019, pp. 4659--4663.

\bibitem{LS_2}
T.~L. Jensen and E.~De~Carvalho, ``An optimal channel estimation scheme for
  intelligent reflecting surfaces based on a minimum variance unbiased
  estimator,'' in \emph{Proc. IEEE Int. Conf. Acoust.,Speech Signal Process.
  (ICASSP)}, 2020, pp. 5000--5004.

\bibitem{MMSE}
Q.-U.-A. Nadeem, H.~Alwazani, A.~Kammoun, A.~Chaaban, M.~Debbah, and M.-S.
  Alouini, ``Intelligent reflecting surface-assisted multi-user {MISO}
  communication: Channel estimation and beamforming design,'' \emph{IEEE Open
  J. Commun. Soc.}, vol.~1, pp. 661--680, 2020.

\bibitem{gui}
G.~Zhou, C.~Pan, H.~Ren, P.~Popovski, and A.~L. Swindlehurst, ``Channel
  estimation for {RIS}-aided multiuser millimeter-wave systems,'' \emph{IEEE
  Trans. Signal Process.}, vol.~70, pp. 1478--1492, 2022.

\bibitem{CS_1}
P.~Wang, J.~Fang, H.~Duan, and H.~Li, ``Compressed channel estimation for
  intelligent reflecting surface-assisted millimeter wave systems,'' \emph{IEEE
  Signal Process. Lett.}, vol.~27, pp. 905--909, 2020.

\bibitem{CS_2}
X.~Wei, D.~Shen, and L.~Dai, ``Channel estimation for {RIS} assisted wireless
  communications—part ii: An improved solution based on double-structured
  sparsity,'' \emph{IEEE Commun. Lett.}, vol.~25, no.~5, pp. 1403--1407, 2021.

\bibitem{RIS_bt1}
C.~You, B.~Zheng, and R.~Zhang, ``Fast beam training for {IRS}-assisted
  multiuser communications,'' \emph{IEEE Wireless Commun. Lett.}, vol.~9,
  no.~11, pp. 1845--1849, 2020.

\bibitem{RIS_bt2}
J.~Zhang, Y.~Huang, J.~Wang, X.~You, and C.~Masouros, ``Intelligent interactive
  beam training for millimeter wave communications,'' \emph{IEEE Trans.
  Wireless Commun.}, vol.~20, no.~3, pp. 2034--2048, 2021.

\bibitem{RIS2}
B.~Ning, Z.~Chen, W.~Chen, Y.~Du, and J.~Fang, ``Terahertz multi-user massive
  {MIMO} with intelligent reflecting surface: Beam training and hybrid
  beamforming,'' \emph{IEEE Trans. Veh. Technol.}, vol.~70, no.~2, pp.
  1376--1393, 2021.

\bibitem{beam_sweep}
T.~Bai and R.~W. Heath, ``Analysis of beam sweep channel estimation in mmwave
  massive {MIMO} networks,'' in \emph{Proc.IEEE GlobalSIP.}, 2016, pp.
  615--619.

\bibitem{h_codebok_1}
Z.~Xiao, T.~He, P.~Xia, and X.-G. Xia, ``Hierarchical codebook design for
  beamforming training in millimeter-wave communication,'' \emph{IEEE Trans.
  Wireless Commun.}, vol.~15, no.~5, pp. 3380--3392, 2016.

\bibitem{h_codebok_2}
Z.~Xiao, H.~Dong, L.~Bai, P.~Xia, and X.-G. Xia, ``Enhanced channel estimation
  and codebook design for millimeter-wave communication,'' \emph{IEEE Trans.
  Veh. Technol.}, vol.~67, no.~10, pp. 9393--9405, 2018.

\bibitem{h_codebok_3}
K.~Chen, C.~Qi, and G.~Y. Li, ``Two-step codeword design for millimeter wave
  massive {MIMO} systems with quantized phase shifters,'' \emph{IEEE Trans.
  Signal Process.}, vol.~68, pp. 170--180, 2020.

\bibitem{jindian}
A.~Alkhateeb, S.~Alex, P.~Varkey, Y.~Li, Q.~Qu, and D.~Tujkovic, ``Deep
  learning coordinated beamforming for highly-mobile millimeter wave systems,''
  \emph{IEEE Access}, vol.~6, pp. 37\,328--37\,348, 2018.

\bibitem{make1}
K.~Ma, D.~He, H.~Sun, Z.~Wang, and S.~Chen, ``Deep learning assisted calibrated
  beam training for millimeter-wave communication systems,'' \emph{IEEE Trans.
  Commun.}, vol.~69, no.~10, pp. 6706--6721, 2021.

\bibitem{make2}
K.~Ma, D.~He, H.~Sun, and Z.~Wang, ``Deep learning assisted mmwave beam
  prediction with prior low-frequency information,'' in \emph{Proc. IEEE Int.
  Conf. Commun.}, 2021, pp. 1--6.

\bibitem{make3}
K.~Ma, P.~Zhao, and Z.~wang, ``Deep learning assisted beam prediction using
  out-of-band information,'' in \emph{Proc. IEEE 91st Veh. Technol. Conf.
  (VTC-Spring)}, 2020, pp. 1--5.

\bibitem{qi}
C.~Qi, Y.~Wang, and G.~Y. Li, ``Deep learning for beam training in millimeter
  wave massive {MIMO} systems,'' \emph{IEEE Trans. Wireless Commun.}, pp. 1--1,
  2020.

\bibitem{RIS_pathloss}
W.~Tang, M.~Z. Chen, X.~Chen, J.~Y. Dai, Y.~Han, M.~Di~Renzo, Y.~Zeng, S.~Jin,
  Q.~Cheng, and T.~J. Cui, ``Wireless communications with reconfigurable
  intelligent surface: Path loss modeling and experimental measurement,''
  \emph{IEEE Trans. Wireless Commun.}, vol.~20, no.~1, pp. 421--439, 2020.

\bibitem{wei_codebook}
X.~Wei, L.~Dai, Y.~Zhao, G.~Yu, and X.~Duan, ``Codebook design and beam
  training for extremely large-scale {RIS}: Far-field or near-field?''
  \emph{China Commun.}, vol.~19, no.~6, pp. 193--204, 2022.

\bibitem{Daill_2}
M.~Cui and L.~Dai, ``Channel estimation for extremely large-scale {MIMO}:
  Far-field or near-field?'' \emph{IEEE Trans. Commun.}, vol.~70, no.~4, pp.
  2663--2677, 2022.

\bibitem{ofdm}
H.~Zhu and J.~Wang, ``Chunk-based resource allocation in {OFDMA} systems—part
  {II}: Joint chunk, power and bit allocation,'' \emph{IEEE Trans. Commun.},
  vol.~60, no.~2, pp. 499--509, 2012.

\bibitem{RIS1}
Q.~Wu and R.~Zhang, ``Intelligent reflecting surface enhanced wireless network
  via joint active and passive beamforming,'' \emph{IEEE Trans. Wireless
  Commun.}, vol.~18, no.~11, pp. 5394--5409, 2019.

\bibitem{RIS3}
Q.-U.-A. Nadeem, H.~Alwazani, A.~Kammoun, A.~Chaaban, M.~Debbah, and M.-S.
  Alouini, ``Intelligent reflecting surface-assisted multi-user {MISO}
  communication: Channel estimation and beamforming design,'' \emph{IEEE Open
  J. Commun. Soc.}, vol.~1, pp. 661--680, 2020.

\bibitem{antenna_sapcing}
K.~T. Selvan and R.~Janaswamy, ``Fraunhofer and fresnel distances : Unified
  derivation for aperture antennas.'' \emph{IEEE Antennas Propag. Mag.},
  vol.~59, no.~4, pp. 12--15, 2017.

\bibitem{near_field1}
Z.~Zhou, X.~Gao, J.~Fang, and Z.~Chen, ``Spherical wave channel and analysis
  for large linear array in los conditions,'' in \emph{Proc. IEEE Globecom
  Workshops (GC Wkshps)}, 2015, pp. 1--6.

\bibitem{steer_vector}
C.~Hu, L.~Dai, T.~Mir, Z.~Gao, and J.~Fang, ``Super-resolution channel
  estimation for mmwave massive {MIMO} with hybrid precoding,'' \emph{IEEE
  Trans. Veh. Technol.}, vol.~67, no.~9, pp. 8954--8958, 2018.

\bibitem{strong_path1}
W.~Roh, J.-Y. Seol, J.~Park, B.~Lee, J.~Lee, Y.~Kim, J.~Cho, K.~Cheun, and
  F.~Aryanfar, ``Millimeter-wave beamforming as an enabling technology for {5G}
  cellular communications: theoretical feasibility and prototype results,''
  \emph{IEEE Commun. Mag.}, vol.~52, no.~2, pp. 106--113, 2014.

\bibitem{strong_path2}
A.~L. Swindlehurst, E.~Ayanoglu, P.~Heydari, and F.~Capolino, ``Millimeter-wave
  massive {MIMO}: the next wireless revolution?'' \emph{IEEE Commun. Mag.},
  vol.~52, no.~9, pp. 56--62, 2014.

\bibitem{near_field_G1}
J.~Sherman, ``Properties of focused apertures in the fresnel region,''
  \emph{IRE Trans. Antennas Propagat.}, vol.~10, no.~4, pp. 399--408, 1962.

\bibitem{near_field_G2}
Y.~Han, S.~Jin, C.-K. Wen, and X.~Ma, ``Channel estimation for extremely
  large-scale massive {MIMO} systems,'' \emph{IEEE Wireless Commun. Lett.},
  vol.~9, no.~5, pp. 633--637, 2020.

\bibitem{wei_beamspace}
X.~Wei, C.~Hu, and L.~Dai, ``Deep learning for beamspace channel estimation in
  millimeter-wave massive {MIMO} systems,'' \emph{IEEE Trans. Commun.},
  vol.~69, no.~1, pp. 182--193, 2021.

\end{thebibliography}


\end{document}